\documentclass[12pt, letterpaper]{article}

\usepackage[letterpaper,margin=1in]{geometry}

\usepackage[ruled,vlined]{algorithm2e}
\usepackage{titlesec}

\renewenvironment{abstract}
	{\quotation}
	{\endquotation}

\date{}

\makeatletter
\renewcommand{\fnum@figure}{\textbf{Figure \thefigure}}
\renewcommand{\fnum@table}{\textbf{Table \thetable}}
\makeatother

\usepackage{scicite}

\usepackage{url}

\usepackage{times}
\usepackage{graphicx,xcolor}
\usepackage{amsmath,amsfonts,amssymb,bm}
\usepackage{listings}
\usepackage{soul}
\usepackage{multirow}
\usepackage{makecell}

\usepackage{setspace}

\usepackage[
unicode=true,
bookmarks=true,%
bookmarksnumbered=true,%
hyperfootnotes=false,
colorlinks=true,%
linkcolor=blue,%
citecolor=blue,%
urlcolor=blue
]{hyperref}  

\newenvironment{mat}{\left[\begin{array}{ccccccccccccccc}}{\end{array}\right]}
\newcommand\bcm{\begin{mat}}
\newcommand\ecm{\end{mat}}

\newenvironment{cmat}{\left(\begin{array}{ccccccccccccccc}}{\end{array}\right)}
\newcommand\bcrm{\begin{cmat}}
\newcommand\ecrm{\end{cmat}}

\newenvironment{rmat}{\left[\begin{array}{rrrrrrrrrrrrr}}{\end{array}\right]}
\newcommand\brm{\begin{rmat}}
\newcommand\erm{\end{rmat}}

\def\scititle{
	Inverse Design of Flat-Foldable Volumetric Origami with Smooth Curved Profile
}
\newcommand{\shorttitle}{Volumetric Origami with Smooth Curvature}
\title{\bfseries \boldmath \scititle}

\author{
	Byoung-Gyu~Kim$^{1}$,
	Geonhee~Cho$^{1}$,\and
	Yasuhiro~Miyazawa$^{1}$,
    Hak-Tae~Lee$^{2}$,
    Jinkyu~Yang$^{1\ast}$\and
	\small$^{1}$Department of Mechanical Engineering, Seoul National University,\and       \small1 Gwanak-ro, Gwanak-gu, Seoul 08826, South Korea\and
	\small$^{2}$Department of Aerospace Engineering and the Program in Aerospace Systems Convergence, Inha University,\and       
    \small100 Inha-ro, Michuhol-gu, Incheon 22212, South Korea\and
	\small$^\ast$Corresponding author. Email: jkyang11@snu.ac.kr
}

\begin{document}

\maketitle

\noindent\textbf{Short title:} \shorttitle

\noindent\textbf{Teaser:} An inverse design transforms curved geometries into flat-foldable volumetric origami while preserving their curved profiles, enabling arbitrarily compact stowage.

\begin{abstract} \bfseries \boldmath
Through flat-folding, origami provides an extremely compact packaging strategy for deployable structures in aerospace, architecture, and robotics.
However, origami’s flat, volumeless facets limit the formation of smooth curvature, restricting its applicability in systems where smooth curved geometries are essential for performance, such as aerospace and electromagnetic communication systems.
Here, we propose volumetric origami that preserves smooth curvature and an inverse design method that generates flat-foldable volumetric origami for given target curved surfaces. 
The flat-foldability enables arbitrarily prescribed compactness in volumetric origami folding, with its stowage efficiency governed by the number of cells and the target profile.
The structural integrity and engineering feasibility of volumetric origami are validated through successful flight testing of a UAV equipped with flat-foldable volumetric origami wings replicating a target airfoil.
Our approach bridges the gap between planar origami and the curvature requirements of engineering systems, expanding design freedom for curved structures under stringent spatial constraints.
\end{abstract}


\section{Introduction}
Origami, an ancient art of paper folding, has emerged as a powerful design tool in diverse fields such as metamaterials~\cite{Jamalimehr_Rigidly_2022,Silverberg_Using_2014,Zhao_Modular_2025,Fang_Modular_2018,Yasuda_Origami_2019}, robotics~\cite{Kim_An_2018,Ze_Soft_2022,Felton_A_2014}, and space exploration~\cite{Miura_Method_1985,Zirbel_Accommodating_2013}, owing to its unique geometry and reconfigurable nature.
In particular, origami-based deployable structures~\cite{Melancon_Multistable_2021,Zhu_Large-scale_2024,Peng_Thick-panel_2025} exhibit the intrinsic capability of achieving highly compact packaging via flat-folding, while maintaining minimal or no facet deformation during deployment.
Unlike bar-linkage mechanisms~\cite{Zhao_The_2009,Mira_Deployable_2014} that cannot form continuous deployed surfaces or inflatable membranes~\cite{Siefert_Bio-inspired_2019,Usevitch_An_2020} that rely on constant pressure to maintain their shape, origami-based deployable structures can be compactly stowed and, once deployed, form self-supporting configurations with continuous, well-defined surfaces.

Despite these distinctive advantages, origami structures are inherently composed of flat facets, which creates a fundamental mismatch with curved geometries. Consequently, addressing stowage requirements for systems involving curvature remains a formidable challenge in deployable structural design. In recent years, considerable efforts have been made to design origami structures capable of approximating or matching curved surfaces upon deployment, using diverse origami patterns such as Miura-ori~\cite{Dudte_2016, Hu_2021, Wang_2016, Wang_2023}, Resch~\cite{Tachi_2013, Yu_2023}, and flasher~\cite{WangS_2022, Tian_2024, Yao_2024}, as well as thick origami~\cite{Zhang_2023, WangC_2022}.
However, these approaches typically produce faceted approximations rather than truly smooth surfaces, as they consist of flat panels that only discretely approximate the target curvature.
This faceted surface limits their applicability in systems that require smooth, continuous curvature for optimal aerodynamic or electromagnetic performance, such as aircraft wings or reflector antennas.
While limited studies~\cite{Tian_2024, Zhang_2023} have realized smooth curvatures, they rely on \textit{planar} origami patterns that are topologically restricted to disk-like surfaces, making them incapable of forming closed or multi-bounded \textit{volumetric} geometries such as generalized cylinders and cones.

In this work, we introduce volumetric origami, a three-dimensional geometric framework that preserves smooth curved profiles upon deployment.
It is composed of spatial panels connected by straight creases, enabling the formation of seamlessly curved surfaces in its fully deployed state.
The inherent three-dimensional nature of volumetric origami overcomes the topological limitations of planar-origami-based structures~\cite{Tian_2024, Zhang_2023}, allowing for the realization of smooth annular surfaces, such as cylinders and cones.
In addition, we establish an inverse design method that converts an arbitrary generalized cylinder or generalized cone into a corresponding flat-foldable volumetric origami, regardless of the cross-sectional shape.
For generalized cylinders with smooth and convex cross sections, we reveal that our inverse design method yields monotonically increasing compactness as the cell count grows, while the polar representation of the cross-sectional profile uniquely determines the per-cell packaging efficiency.
Finally, we demonstrate the practical applicability of our approach by designing and fabricating a deployable origami wing, whose structural integrity and aerodynamic performance are validated through successful flight tests.

\section{Results}
\subsection{Volumetric origami}

\begin{figure*}[t]
    \centering
    \includegraphics[width=184mm]{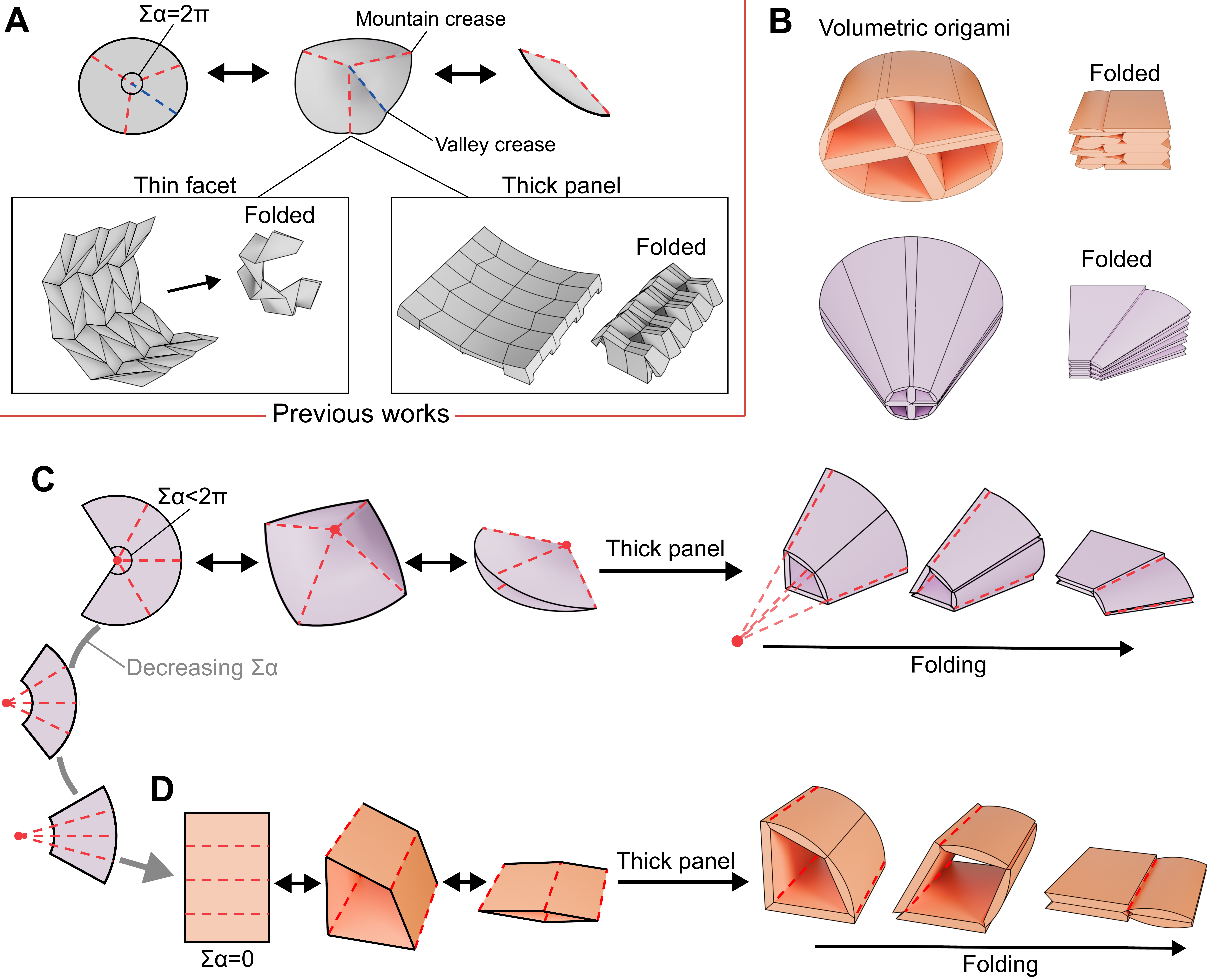}
\caption{
\textbf{Concept of flat-foldable volumetric origami with a smooth curved surface.} 
(\textbf{A}) Origami targeting curved surfaces based on developable and flat-foldable patterns (e.g., Miura-ori). Left: thin-facet construction. Right: thick-panel construction with smooth curved surfaces.
(\textbf{B}) Two representative examples of flat-foldable volumetric origami. Top: cylindrical deployed shape. Bottom: conical deployed shape.
(\textbf{C}) Flat-foldable volumetric origami cell constructed from non-developable origami (sum of sector angles $< 2\pi$).
(\textbf{D}) Flat-foldable volumetric origami cell constructed from a special case of a non-developable origami pattern (sum of sector angles $= 0$). }
    \label{intro_fig}
\end{figure*}

Previous origami designs for approximating or matching curved surfaces have predominantly relied on developable crease patterns (Fig.~\ref{intro_fig}A). In these designs, the sector angles around each vertex sum to $2\pi$, allowing the crease patterns to be mapped onto a two-dimensional plane, regardless of whether they employ thin facets~\cite{Dudte_2016, Hu_2021, Wang_2016, Wang_2023, Tachi_2013, Yu_2023, WangS_2022, Tian_2024, Yao_2024} or thick panels~\cite{Zhang_2023, WangC_2022}. Because these approaches are constrained by planar topology under the condition of rigid foldability, they are inherently limited to disk-like surfaces, such as parabolic antennas. However, many essential engineering geometries possess non-planar topologies (e.g., cylinders and cones in  Fig.~\ref{intro_fig}B), necessitating a folding strategy that transcends two-dimensional constraints. To accommodate such surfaces, the crease pattern must depart from a planar configuration so that the origami’s topology can deviate from a planar domain.

Accordingly, we initiate our design from a single-degree-of-freedom, non-developable origami pattern characterized by a sum of sector angles less than $2\pi$ at the vertex. As illustrated in Fig.~\ref{intro_fig}C, the crease network of this pattern is intrinsically embedded in three-dimensional space. To achieve smooth curvature, we replace conventional flat facets with spatial panels featuring smooth curved surfaces. We define the resulting structure as a volumetric origami cell, which maintains its foldability while preserving a continuous curved profile. Fig.~\ref{intro_fig}D illustrates a specific case of this cell where the angular deficit is maximized such that the sum of sector angles becomes zero. Notably, the volumetric origami cell shown in Fig.~\ref{intro_fig}C serves as the building block for conical structures, whereas the configuration in Fig.~\ref{intro_fig}D leads to cylindrical (or prismatic) volumetric origami. By assembling these cells, we can construct volumetric origami capable of forming three-dimensional architecture like cylinders and cones (Fig.~\ref{intro_fig}B).

\subsection{Inverse design of flat-foldable volumetric origami}
For the volumetric origami introduced in Fig.~\ref{intro_fig} to be effectively utilized in engineering applications, such as aerodynamic wings or high-gain antennas, its surface geometry must be accurately defined to achieve targeted performance metrics. To this end, we develop an inverse design framework that systematically maps a prescribed target surface (Input, Fig.~\ref{cylinder_inverse}A) to its corresponding flat-foldable volumetric origami configuration (Output, Fig.~\ref{cylinder_inverse}A). The requirement of flat-foldability is strictly enforced to ensure maximum stowage efficiency. Consistent with the geometric definitions in Fig.~\ref{intro_fig}D, we assume the target surface is a generalized cylinder, with its cross-sectional profile in the $xy$-plane denoted by $\mathbf{f}(s) = x(s)\mathbf{i} + y(s)\mathbf{j}$ (orange curve, Fig.~\ref{cylinder_inverse}A). 

The topological backbone of the resulting volumetric origami is defined as a planar graph composed entirely of quadrilateral internal faces. In this representation, edges correspond to panels, vertices correspond to junctions, and internal faces correspond to individual volumetric origami cells. We refer to such planar graphs as origami layouts and generate the complete set of valid layouts using a canonical augmentation algorithm (see Supplementary Note 1 for details). A candidate origami layout is then selected and embedded in the $xy$-plane to establish its initial geometric realization, providing the structural foundation for the inverse design process.

Starting from this initial geometric description, each internal face is transformed into a flat-foldable volumetric origami cell, consistent with the configuration shown in Fig.~\ref{intro_fig}D. As illustrated in stage (I) of Fig.~\ref{cylinder_inverse}A, given the vertex coordinates $\mathbf{q}_1, \mathbf{q}_2, \mathbf{q}_3, \mathbf{q}_4$ and the target cross-section $\mathbf{f}(s)$, we determine the precise locations of the creases and the corresponding panel geometries requisite for flat-folding. First, we position creases $\mathbf{q}_2'$ and $\mathbf{q}_4'$ on the target profile $\mathbf{f}(s)$ such that $\mathbf{q}_2' = \mathbf{f}(s_2')$ and $\mathbf{q}_4' = \mathbf{f}(s_4')$. Crucially, these locations are determined to ensure that the resulting volumetric origami can be flat-folded without geometric self-intersection. To achieve this, we stipulate that every volumetric origami cell maintains a constant panel thickness (Fig.~S4B), facilitating a collision-free stacked state (Fig.~S4E). Consequently, with the thicknesses of the thick panels defined as $T_1(s_2')$, $T_2(s_2')$, $T_3(s_4')$, and $T_4(s_4')$ (stage (II), Fig.~\ref{cylinder_inverse}A), the parameters $s_2'$ and $s_4'$ are governed by the thickness-dependent flat-foldability constraints:
\begin{equation}\label{eq:cylinder_thickness_condition1}
T_1(s_2') = T_2(s_2'),
\end{equation}
\begin{equation}\label{eq:cylinder_thickness_condition2}
T_3(s_4') = T_4(s_4').
\end{equation}
Detailed expressions and derivations for Eqs.~\eqref{eq:cylinder_thickness_condition1} and~\eqref{eq:cylinder_thickness_condition2} are provided in Supplementary Note 2. 

By locating creases $\mathbf{v}_1$ and $\mathbf{v}_3$ as shown in stage (III), we construct a cross-section that satisfies the thickness constraints. 
However, satisfying these local constraints does not inherently guarantee global flat-folding, as exemplified by the kinematic incompatibility shown in (III-i). 
To ensure global flat-foldability, we adjust the positions of $\mathbf{v}_1$ and $\mathbf{v}_3$ by translating the auxiliary lines inward by $t_1$ and $t_2$ (stage (IV), Fig.~\ref{cylinder_inverse}A). These translation distances are applied in a manner that preserves the thickness constraints. To reach a fully compatible flat-folded state, $t_1$ and $t_2$ must satisfy the lengthwise flat-foldability constraint:
\begin{equation}\label{eq:cylinder_length_condition}
l_1(t_1,t_2) = l_2(t_1,t_2),
\end{equation}
where $l_1(t_1,t_2)$ and $l_2(t_1,t_2)$ represent the distances between $\mathbf{v}_1$ and $\mathbf{v}_3$ when the folding angles at $\mathbf{q}_2'$ and $\mathbf{q}_4'$ are fully actuated to $180^\circ$ (stage (IV-i)). Eq.~\eqref{eq:cylinder_length_condition} ensures that the upper and lower panel pairs converge at coincident creases $\mathbf{v}_1$ and $\mathbf{v}_3$, implying that the interior folding angles vanish (i.e., reach $0^\circ$) when the primary creases are fully folded. Finally, as shown in stage (V), the flat-foldable volumetric origami cell is realized by extruding the stage (IV) configuration along the $z$-axis.

Following the procedure described above, every internal face of the embedded origami layout is transformed into a flat-foldable volumetric origami cell, as shown in stage (VI) of Fig.~\ref{cylinder_inverse}A. Subsequently, these individual cells are integrated by bonding them along the internal edges of the layout, thereby completing the inverse design and realizing the final flat-foldable volumetric origami. Crucially, to preserve the system's mobility, one internal edge incident to each internal vertex is intentionally left unbonded (see Supplementary Note 2 for details on bonding strategies and their impact on kinematics). In contrast, interlocking all internal edges introduces over-constraining, resulting in a mechanically locked state that transforms the assembly into a load-bearing structure. This structural characteristic is strategically exploited in a subsequent section to enhance structural rigidity of a deployable wing under aerodynamic loads.

As the proposed framework is independent of cross-sectional geometry, the inverse design approach is applicable to arbitrary generalized cylinders. For instance, as shown in Fig.~\ref{cylinder_inverse}B,C, the method successfully accommodates complex profiles that include both concave and convex regions, such as the clover-shaped cross-section. Moreover, as demonstrated in Fig.~\ref{cylinder_inverse}D,E, the robustness of the algorithm allows for the inclusion of cusp points, as exemplified by the heart-shaped profile. The continuous folding kinetics of the physical prototypes shown in Fig.~\ref{cylinder_inverse}B--E are documented in Movie~S1.

\subsection{Extension to generalized cones}
Building upon the inverse design framework established for generalized cylinders, we extend our methodology to encompass generalized cones. A generalized cone is defined as a surface generated by the linear expansion of an arbitrary two-dimensional closed curve from an apex $\mathbf{o}$, as illustrated in stage (I) of Fig.~\ref{Cone_inverse}A. For these conical geometries, we execute the inverse design on a spherical domain (gray surface, Fig.~\ref{Cone_inverse}A), rather than the $xy$-plane employed for cylindrical cases. This shift in the design domain is necessitated by the kinematic behavior of conical origami. In a generalized cone, planar-based inverse design results in creases that are oblique to the design plane; consequently, the geometry deviates from the plane during folding, failing to provide a consistent two-dimensional representation of the full three-dimensional folding motion (see Fig. S8). In contrast, performing the inverse design on a spherical surface ensures that all crease lines remain normal to the sphere's surface. Since spherical figures are preserved under rotations about axes passing through the center $\mathbf{o}$, this spherical representation guarantees that the volumetric origami remains intrinsically mapped to the sphere throughout its entire folding range. This strategic projection allows the complex three-dimensional kinematics of conical volumetric origami to be fully captured on a two-dimensional spherical manifold, enabling the seamless adaptation of our cylinder-based inverse design method to generalized cones.

Following the logic applied to generalized cylinders, we select a generated origami layout and embed it to achieve geometric realization; here, however, this embedding is performed on a spherical manifold rather than a Euclidean plane ((II), Fig.~\ref{Cone_inverse}A). Subsequently, each internal face of the layout embedded on the spherical domain is converted into a volumetric origami cell with a smooth curvilinear outer surface, as shown in Fig.~\ref{Cone_inverse}B. Starting from the layout vertices $\mathbf{p}_1, \mathbf{p}_2, \mathbf{p}_3$, and $\mathbf{p}_4$, our objective is to determine the precise locations of the creases $\mathbf{p}'_2, \mathbf{p}'_4, \mathbf{u}_1$, and $\mathbf{u}_3$ to ensure flat-foldability. First, as shown in stage (I) of Fig.~\ref{Cone_inverse}B, we position the creases $\mathbf{p}_2'=\mathbf{g}(s_2')$ and $\mathbf{p}_4'=\mathbf{g}(s_4')$ on the target profile $\mathbf{g}(s)$, defined as the intersection between the generalized cone and the spherical design surface (red curve, Fig.~\ref{Cone_inverse}A). The positions of these creases are determined by solving the following non-penetration equations for $s_2'$ and $s_4'$:
\begin{equation}
\label{eq:cone_thickness_condition1}
\min_{s_3 \le s \le s_2'} \, \mathbf{R}(\mathbf{g}(s_2'), \phi_{\mathbf{g}(s_2')}) \mathbf{g}(s) \cdot (\mathbf{p}_2 \times \mathbf{p}_1)
= 0,
\end{equation}
\begin{equation}
\label{eq:cone_thickness_condition2}
\min_{s_4' \le s \le s_3} \, \mathbf{R}(-\mathbf{g}(s_4'), \phi_{\mathbf{g}(s_4')}) \mathbf{g}(s) \cdot (\mathbf{p}_1 \times \mathbf{p}_4)
= 0,
\end{equation}
where $\mathbf{R}(\mathbf{a}, b)$ denotes the rotation matrix derived by Rodrigues' formula for a rotation about axis $\mathbf{a}$ by angle $b$, and $\phi_{\mathbf{g}(s_2')}$ is the prescribed rotation angle corresponding to a fully folded state (folding angle = $180^\circ$). Eqs.~\eqref{eq:cone_thickness_condition1} and \eqref{eq:cone_thickness_condition2} are the spherical analogues of the thickness constraints in Eqs.~\eqref{eq:cylinder_thickness_condition1} and \eqref{eq:cylinder_thickness_condition2}, formulated to prevent geometric interference. Specifically, these constraints ensure that the panels adjacent to $\mathbf{p}'_2$ and $\mathbf{p}'_4$ converge onto respective great circles without overlapping in the folded configuration.

Next, as shown in stage (III) of Fig.~\ref{Cone_inverse}B, the internal arcs are rotated inward by angles $\theta_1$ and $\theta_2$ to form internal spatial facets. The locations of the creases $\mathbf{u}_1$ and $\mathbf{u}_3$ are determined by the intersection of these rotated arcs, with the angles $\theta_1$ and $\theta_2$ selected to satisfy the following kinematic constraint:
\begin{equation}\label{eq:cone_length_condition}\mathbf{R}(\mathbf{p}_2', \phi_{\mathbf{p}_2'}) \mathbf{u}_3(\theta_1, \theta_2) \cdot \mathbf{u}_1(\theta_1, \theta_2) = \mathbf{R}(-\mathbf{p}_4', \phi_{\mathbf{p}_4'}) \mathbf{u}_3(\theta_1, \theta_2) \cdot \mathbf{u}_1(\theta_1, \theta_2).
\end{equation}
Eq.~\eqref{eq:cone_length_condition} represents the spherical-geometric counterpart of the lengthwise constraint in Eq.~\eqref{eq:cylinder_length_condition}, replacing Euclidean distances with arc lengths on the unit sphere. This constraint ensures that the crease network can undergo large-deformation folding without inducing structural strain. Finally, as shown in stage (IV) of Fig.~\ref{Cone_inverse}B, the spherical geometry is radially extruded from the apex $\mathbf{o}$, generating the final three-dimensional volumetric origami cell and its associated spatial crease lines.

As shown in Fig.~\ref{Cone_inverse}C,E, the final volumetric origami structure is assembled following the same principles as the cylindrical case, by bonding the flat-foldable volumetric origami cells along the internal edges of the spherical layout. Consistent with our previous findings, the inverse design framework for generalized cones remains agnostic to the cross-sectional profile. This robustness allows for the generation of diverse conical geometries, as demonstrated by the circular and elliptical prototypes shown in Fig.~\ref{Cone_inverse}C--F. The continuous folding kinematics of these physical prototypes are demonstrated in Movie~S1.

\subsection{Analysis of packaging efficiency}\label{sec:onesoliton}

In the inverse design of both generalized cylinders and cones, the flat-foldability requirement is strictly enforced to facilitate a compact stowed configuration. Then, to what extent can flat-foldability guarantee compactness in the folded state? To quantitatively assess the degree of compactness, we introduce the contraction ratio, $r_c$, defined as the ratio of the total cross-sectional area of the volumetric origami in its flat-folded state ($a$) to that of the target shape ($A$), expressed as $r_c = a/A$. As illustrated in Fig.~\ref{PR_analysis}A, the folded area $a$ is calculated as the sum of the individual bounded cross-sectional areas of the constituent cells, $a_i$. Thus, the total contraction ratio is given by $r_c = \sum a_i / A$, providing a direct measure of stowage efficiency where a lower value indicates a more significant volume reduction relative to the deployed state.

The contraction ratio is intrinsically linked to the geometry of the volumetric origami, which is governed by the target shape and the layout embedding (i.e., the specific coordinates of the vertices in the chosen origami layout). Consequently, the contraction ratio $r_c$ can be minimized by numerically optimizing the vertex positions. Utilizing the optimization framework detailed in Supplementary Note 4, we observe that increasing the number of volumetric origami cells $n$ yields a progressively lower optimized contraction ratio, as illustrated in Fig.~\ref{PR_analysis}B. 

To systematically investigate the relationship between the optimized contraction ratio and the cell count, we first constrain the target shape to a generalized cylinder with a convex and differentiable cross-section. Under this assumption, we consider an origami layout characterized by a single internal vertex as a baseline for our analysis. 
We further assume that the central angle of each internal face, $\Delta \theta_i$ (Fig.~\ref{PR_analysis}A), is sufficiently small. This assumption allows the flat-foldability constraints in Eqs.~\eqref{eq:cylinder_thickness_condition1}, \eqref{eq:cylinder_thickness_condition2},and~\eqref{eq:cylinder_length_condition} to be treated analytically, enabling an estimation of $a_i$. Following the derivation in Supplementary Note 4, the contraction ratio can be expressed as:
\begin{equation}
\label{eq:discretized_contraction_ratio}
r_c \simeq \frac{1}{A} \sum_{i=1}^{n} \frac{1}{16} r(\theta_i) |\kappa(\theta_i)| \left( r(\theta_i)^2 + 2r'(\theta_i)^2 \right) \Delta \theta_i^2,
\end{equation}
where $r(\theta)$ is the radial distance from the internal vertex (Fig.~\ref{PR_analysis}A), $r'(\theta_i) = \left. dr/d\theta \right|_{\theta_i}$, and $\kappa(\theta_i)$ denotes the curvature of the profile at $\theta_i$. Here, we assume the translation distances $t_1$ and $t_2$ are minimized to maximize compactness.

To determine the global minimum of Eq.~\eqref{eq:discretized_contraction_ratio}, we apply the Cauchy--Schwarz inequality, which provides a rigorous lower bound. This bound is physically attainable when the folded areas are uniformly distributed across all cells (i.e., $a_1 = \dots = a_n$). Consequently, in the continuum limit where $\Delta \theta_i$ vanishes as the number of cells $n$ increases, the optimized contraction ratio $r_{c|\mathrm{opt}}$ can be expressed as:
\begin{equation}\label{eq:continuum_opt_contraction_ratio}
r_{c|\mathrm{opt}} \simeq \frac{1}{\eta} \cdot \frac{1}{n}.
\end{equation}
In this expression, $\eta$ represents a shape-dependent, per-cell packaging efficiency that accounts for the target geometry and the position of the internal vertex:
\begin{equation}
\label{eq:per_cell_efficiency}
\eta = \frac{A}{\left( \int_{0}^{2\pi} \sqrt{\frac{1}{16} r(\theta) |\kappa(\theta)| \left( r(\theta)^2 + 2r'(\theta)^2 \right)} d\theta \right)^2}.
\end{equation}
Detailed derivations and the transition from the discrete summation to the integral form are provided in Supplementary Note 4.

As shown in Fig.~\ref{PR_analysis}C and Fig.~S12, the numerically optimized contraction ratios for various target shapes asymptotically approach the analytical prediction derived in Eq.~\eqref{eq:continuum_opt_contraction_ratio} as the number of cells increases. Furthermore, as illustrated in Fig.~S12B, the distribution of $a_i$ becomes increasingly uniform in the large-$n$ regime. This convergence confirms that the numerical optima coincide with the theoretical condition for the Cauchy–Schwarz equality ($a_1 = \dots = a_n$), thereby validating the continuum-limit approximation in Eq.~\eqref{eq:continuum_opt_contraction_ratio}. Eq.~\eqref{eq:continuum_opt_contraction_ratio} reveals a fundamental scaling law: the optimal contraction ratio is inversely proportional to the number of cells. This implies that for convex and smooth generalized cylinders, the proposed inverse design framework can achieve arbitrarily high compactness, limited only by the chosen discretization density, simply by increasing the cell count. However, in practical applications, the number of cells cannot be increased indefinitely due to physical constraints. As the cell count increases, the cumulative volume occupied by the hinges and the finite thickness of the panels, which are neglected in the idealized model, begin to dominate the folded state, eventually counteracting the benefits of further discretization. Therefore, to achieve superior compactness with a limited number of cells, it is crucial to maximize the per-cell packaging efficiency, $\eta$.

The per-cell packaging efficiency is intrinsically dictated by the target geometry and the spatial positioning of the internal vertex. Fig.~\ref{PR_analysis}D illustrates the dependence of $\eta$ on the target cross-section, modeled as a superellipse:$$x=\operatorname{sgn}(\cos\theta)|\cos\theta|^{2/m}, \quad y=\gamma\,\mathrm{sgn}(\sin\theta)|\sin\theta|^{2/m}$$where the internal vertex is fixed at the origin. Our analysis shows that relative to a circular cross-section ($m=2, \gamma=1$), the per-cell packaging efficiency tends to increase as the shape becomes more polygonal ($m > 2$), where the curvature is concentrated at the corners and nearly zero over most regions. Conversely, $\eta$ decreases as the aspect ratio $\gamma$ deviates from unity, as seen in elongated elliptical shapes where the increased perimeter-to-area ratio complicates efficient spatial packing.

\subsection{Deployable origami wing and flight test}\label{sec:twosoliton}

To demonstrate the practicality of our inverse design framework, we present a deployable aircraft wing in Fig.~\ref{d_wing}. In this design, we configured the airfoil with multiple volumetric cells to enhance packaging efficiency while ensuring the layout was strategically embedded so that the internal cell walls (i.e., spars) remain vertical to take shear loads effectively (Fig.~\ref{d_wing}A,B). Based on this approach, we fabricated a prototype with the NACA 2412 airfoil profile using lightweight polylactic acid (LW-PLA), as shown in Fig.~\ref{d_wing}C (see Methods and also Movie~S2 for folding kinematics). The resulting design achieves a high level of compactness, with a calculated contraction ratio ($r_c$) of approximately 48.3\%, effectively reducing its chord and volume significantly in the flat-folded state. This is followed by a secondary spanwise zig-zag folding that further compresses the span in proportion to the number of segments. To ensure operational reliability, the internal spars were reinforced with carbon fiber-reinforced plastic (CFRP) tubes. Moreover, the fully deployed state is securely maintained through a combination of interference fits and magnetic connectors, providing the necessary structural rigidity to withstand aerodynamic loads during flight. As shown in Fig. S13 and Supplementary Note 5, the prototype successfully passed pre-flight loading tests, demonstrating the required load-carrying capability of our volumetric origami-based wing system.

To demonstrate the practical feasibility of our approach, we integrated the deployable origami wing into a custom-built UAV (Fig.~\ref{d_wing}D), designed in accordance with the specifications in Methods and Supplementary Note 5. The UAV, operated via ground-based remote control, successfully performed a stable flight, including controlled altitude adjustments and coordinated turning maneuvers (Fig.~\ref{d_wing}E and Movie~S2). The successful flight trial confirms that our inverse design framework accurately reproduces the target NACA 2412 airfoil profile, generating sufficient aerodynamic lift while maintaining structural integrity under operational loads. This result highlights the potential of volumetric origami in developing high-performance, deployable aerodynamic surfaces.

\section{Discussion}
We have developed a geometric framework that enables the accurate realization of smooth curvilinear surfaces while strictly preserving flat-foldable kinematics. For generalized cylinders, we define and impose these constraints in Euclidean geometric form directly within the inverse design process. For generalized cones, the constraints are reformulated in a spherical geometric form, reducing the problem dimensionality from three dimensions to two and significantly simplifying the solution. This approach uncovers a fundamental scaling law, a direct trade-off between discretization density and packaging efficiency, that establishes a theoretical guideline for the achievable compactness of ruled surface origami.

This methodology provides a deterministic pathway to program complex curvatures into foldable matter. Such a capability is important for architected materials and autonomous systems where maintaining a precise aerodynamic or electromagnetic profile is essential. By encoding this prescribed curvature into a single degree of freedom system, we can potentially eliminate the need for complex actuation, pointing toward a new class of self-deploying structures that rely on their own geometric constraints for functional reliability.

Looking ahead, these findings open several promising directions. While we have focused on ruled surfaces, extending these inverse design principles to doubly curved geometries, such as hyperbolic or elliptical paraboloids, or even non-orientable structures like M\"obius strips, presents an exciting challenge that will likely necessitate non-Euclidean tessellations. Furthermore, integrating active materials or multi-stability into these volumetric cells could lead to intelligent surfaces capable of adaptive morphing in response to environmental stimuli. Ultimately, combining this geometric versatility with advanced additive manufacturing will pave the way for a new generation of high-performance deployable systems, ranging from micro medical robots to planetary-scale habitats.



\section*{Materials and Methods}
\subsection*{Flat-foldable volumetric origami prototype fabrication}
The physical prototypes presented in Fig.~\ref{cylinder_inverse} and Fig.~\ref{Cone_inverse} were fabricated via fused deposition modeling (FDM) using a Prusa MK4 3D printer (Prusa Research). These structures were printed with polylactic acid filament (PLA-Basic, eSUN). For the deployable wing shown in Fig.~\ref{d_wing}, we employed a lightweight polylactic acid (ePLA-LW, eSUN) to minimize structural weight.
The origami creases were constructed using either 0.2 mm-thick polyethylene terephthalate (PET) sheets or a specialized polyester film (Oracover, Lanitz-Prena Folien Factory GmbH). These flexible membranes were securely bonded to the 3D printed rigid panels using an ethyl cyanoacrylate adhesive (Loctite 401, Henkel), ensuring robust hinge rotation and structural integrity during the folding cycles.

\subsection*{Flight demonstration of the prototype UAV}
To validate the engineering feasibility of our design, we developed a demonstrator aircraft integrated with the previously fabricated foldable wing. Given that each prototype wing weighed 260 g, we prioritized minimizing the mass of other airframe components to maintain a total takeoff weight comparable to similar-scale unmanned aircraft. Since the folding mechanism precluded the installation of onboard ailerons, we incorporated center wing mounts with a 10° dihedral angle to ensure passive roll stability. These mounts also served a structural role, transferring bending loads through a carbon fiber tube integrated into the upper main spar. The addition of these center mounts extended the total wingspan by 60 mm. The aerodynamic configuration was analyzed using Athena Vortex Lattice (AVL)~\cite{Drela_AVL}, with the resulting aircraft specifications detailed in Table~\ref{tab:aircraft_spec}. We estimated the drag of the wing and horizontal tail as a function of the lift coefficient $C_L$ by deriving sectional drag coefficients $C_d$ from two-dimensional airfoil data, specifically, the NACA 2412 profile for the wing and a flat plate model for the tail. The drag contributions from the fuselage and vertical stabilizer were calculated using skin friction coefficients and standard form factors. The resulting drag polar is characterized by the relationship $C_D = 0.029 + 0.072 (C_L - 0.034)^2$. Finally, the motor and propeller were selected by applying the electric propulsion system analysis framework described in~\cite{Nam_Electric_2025}. See Supplementary Note 5 for more details.

\clearpage


\begin{table}[t]
    \centering
    \caption{Aircraft specifications.}
    \label{tab:aircraft_spec}
    \begin{tabular}{l c}
        \hline
        Takeoff mass & 850 g \\
        Wingspan & 1.26 m \\
        Wing area & 0.315 m$^2$ \\
        Cruise speed & 8.5 m/s \\
        Cruise $C_L$ & 0.65 \\
        Static margin & 8--10\% \\
        Motor & T-Motor AS2304 KV1800 \\
        Propeller diameter & 0.203 m \\
        Electric propulsion system efficiency & 45\% \\
        Estimated flight time & 9 minutes \\
        \hline
    \end{tabular}
\end{table}

\clearpage


\clearpage
\newgeometry{left=0.75in,right=0.75in,top=1in,bottom=1in}

\begin{figure*}[t]
    \centering
    \includegraphics[width=150mm]{main_fig/cylinder_inverse.pdf}

\caption{
\textbf{Inverse design of flat-foldable volumetric origami for generalized cylinders.} 
(\textbf{A}) An inverse design process that generates a flat-foldable volumetric origami from a given target shape. Blue dots indicate the locations of vertices in the origami layout, while purple dots indicate the locations of creases on the $xy$-plane.
(\textbf{B}) Inverse design of a flat-foldable volumetric origami targeting a cylinder with a clover-shaped cross section.
(\textbf{C}) Folding motion of physical prototype fabricated based on the design obtained in (B).
(\textbf{D}) Inverse design of a flat-foldable volumetric origami targeting a cylinder with a heart-shaped cross section.
(\textbf{E}) Folding motion of physical prototype fabricated based on the design obtained in (D).}

    \label{cylinder_inverse}
\end{figure*}

\begin{figure*}[t]
    \centering
    \includegraphics[width=184mm]{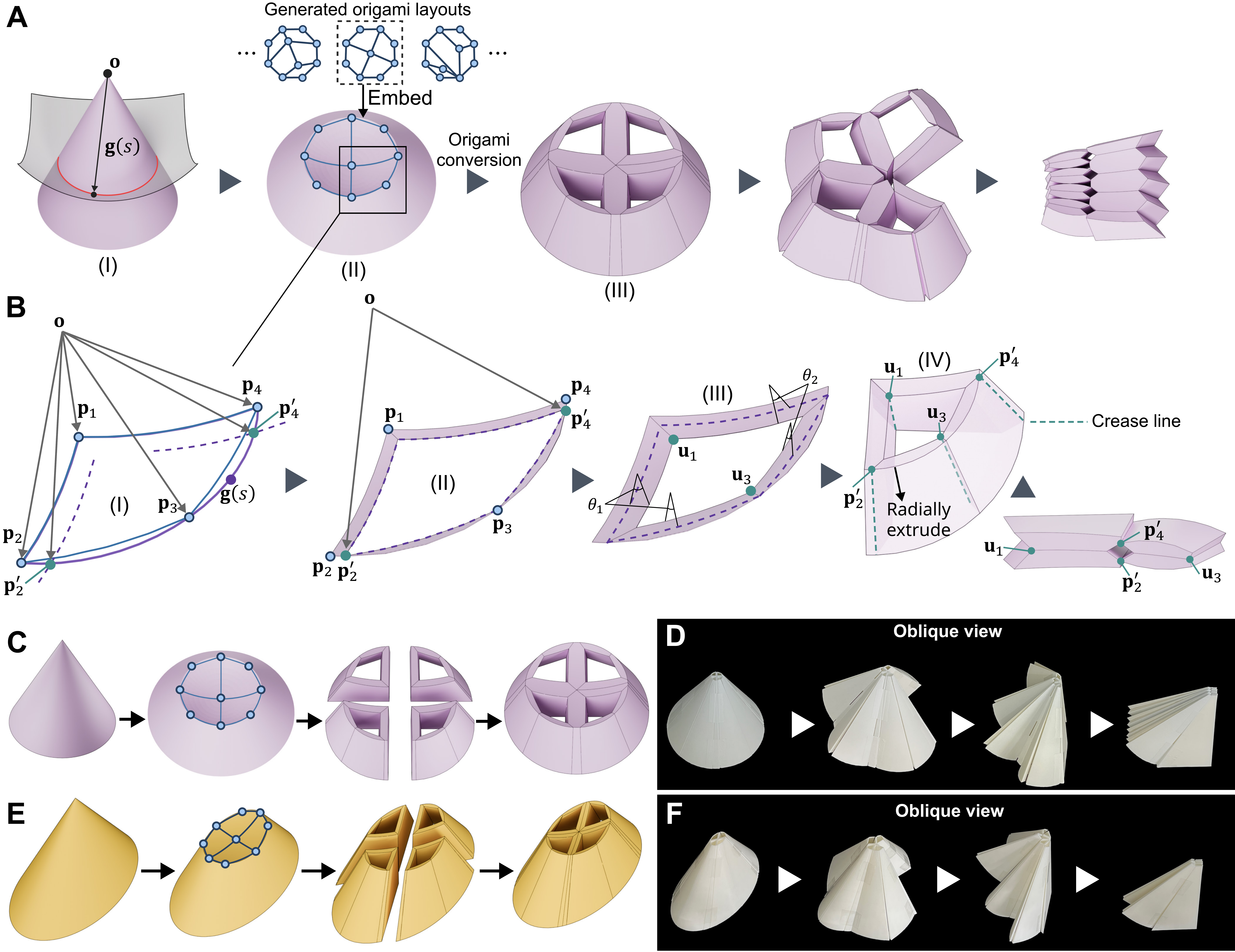}
\caption{
\textbf{Inverse design of flat-foldable volumetric origami for generalized cones.} 
(\textbf{A}) Overall inverse design process for flat-foldable volumetric origami of a generalized cone. The inverse design process is performed on a unit sphere centered at the point $\mathbf{o}$, shown as the gray surface. 
(\textbf{B}) Detailed inverse design process that converts a given internal face of the embedded origami layout with four blue vertices ($\mathbf{p}_1, \mathbf{p}_2, \mathbf{p}_3, \mathbf{p}_4$) into a corresponding flat-foldable volumetric origami cell with four teal creases ($\mathbf{p}_2', \mathbf{p}_4', \mathbf{u}_1, \mathbf{u}_3$). 
(\textbf{C}) Inverse design of a flat-foldable volumetric origami targeting a cone with a circular cross section.
(\textbf{D}) Folding motion of physical prototype fabricated based on the design obtained in (C).
(\textbf{E}) Inverse design of a flat-foldable volumetric origami targeting a cone with an elliptical cross section.
(\textbf{F}) Folding motion of physical prototype fabricated based on the design obtained in (E).
}
    \label{Cone_inverse}
\end{figure*}

\begin{figure*}[t]
    \centering
    \includegraphics[width=184mm]{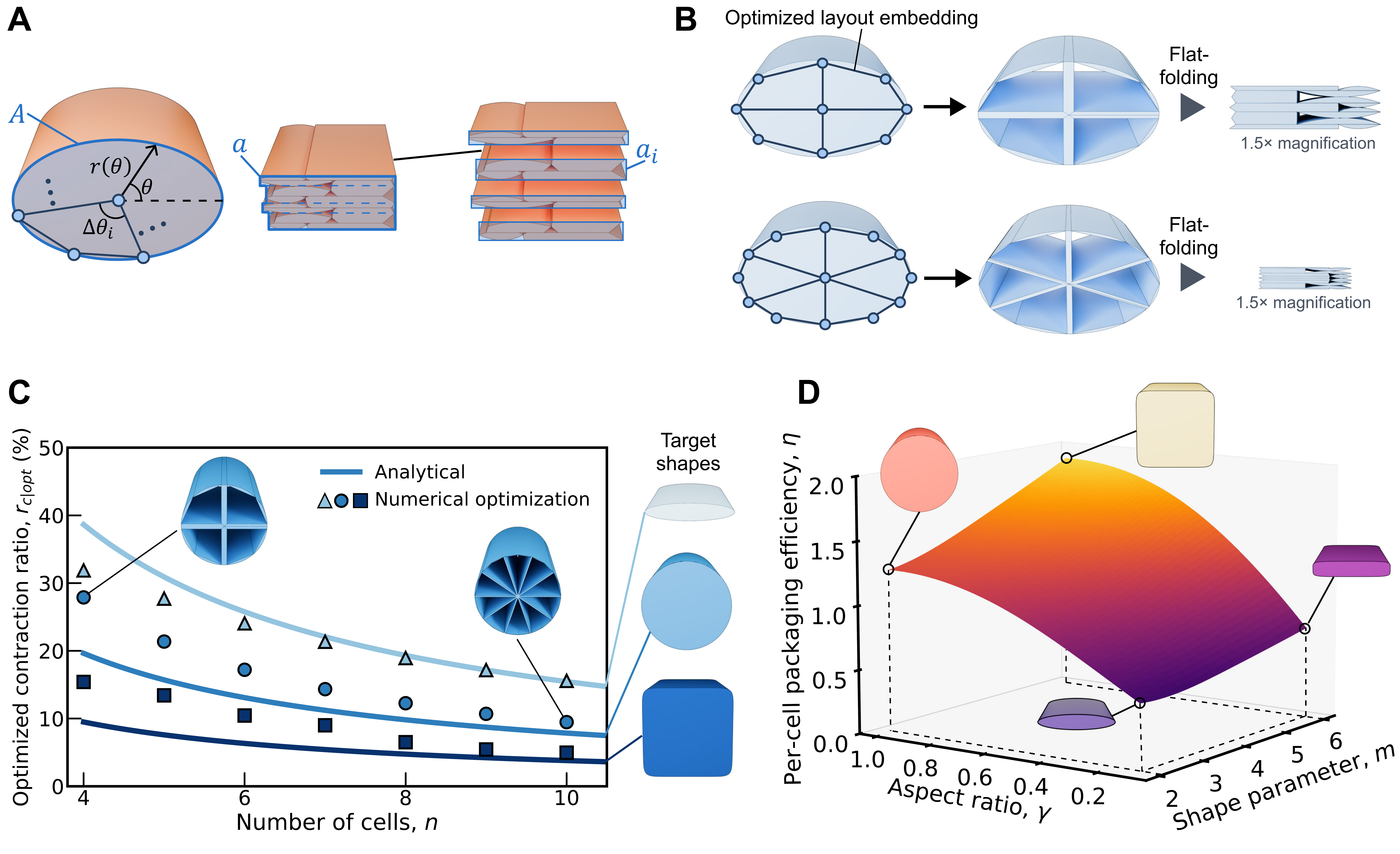}
\caption{
\textbf{Relationship between optimal contraction ratio and the number of cells.} 
(\textbf{A}) Definition of the cross-sectional areas for the contraction ratio.
(\textbf{B}) Conversion of a generalized cylinder with an elliptical cross-section (eccentricity 0.8) into flat-foldable volumetric origami structures based on layouts with 4 and 6 internal faces (4-cell and 6-cell cases). The positions of layout vertices are numerically optimized during graph embedding to achieve lower contraction ratios, yielding values of 24\% and 16\%, respectively.
(\textbf{C}) Plot of the optimized contraction ratio versus the number of cells $n$ for generalized cylinders with various cross-sections. The solid lines represent the analytical continuum-limit expression of the optimal contraction ratio, while the markers indicate local minima obtained through numerical optimization. 
(\textbf{D}) 3D surface plot of the analytically obtained per-cell packaging efficiency $\eta$ for superellipses.}
    \label{PR_analysis}
\end{figure*}

\begin{figure*}[t]
    \centering
    \includegraphics[width=120mm]{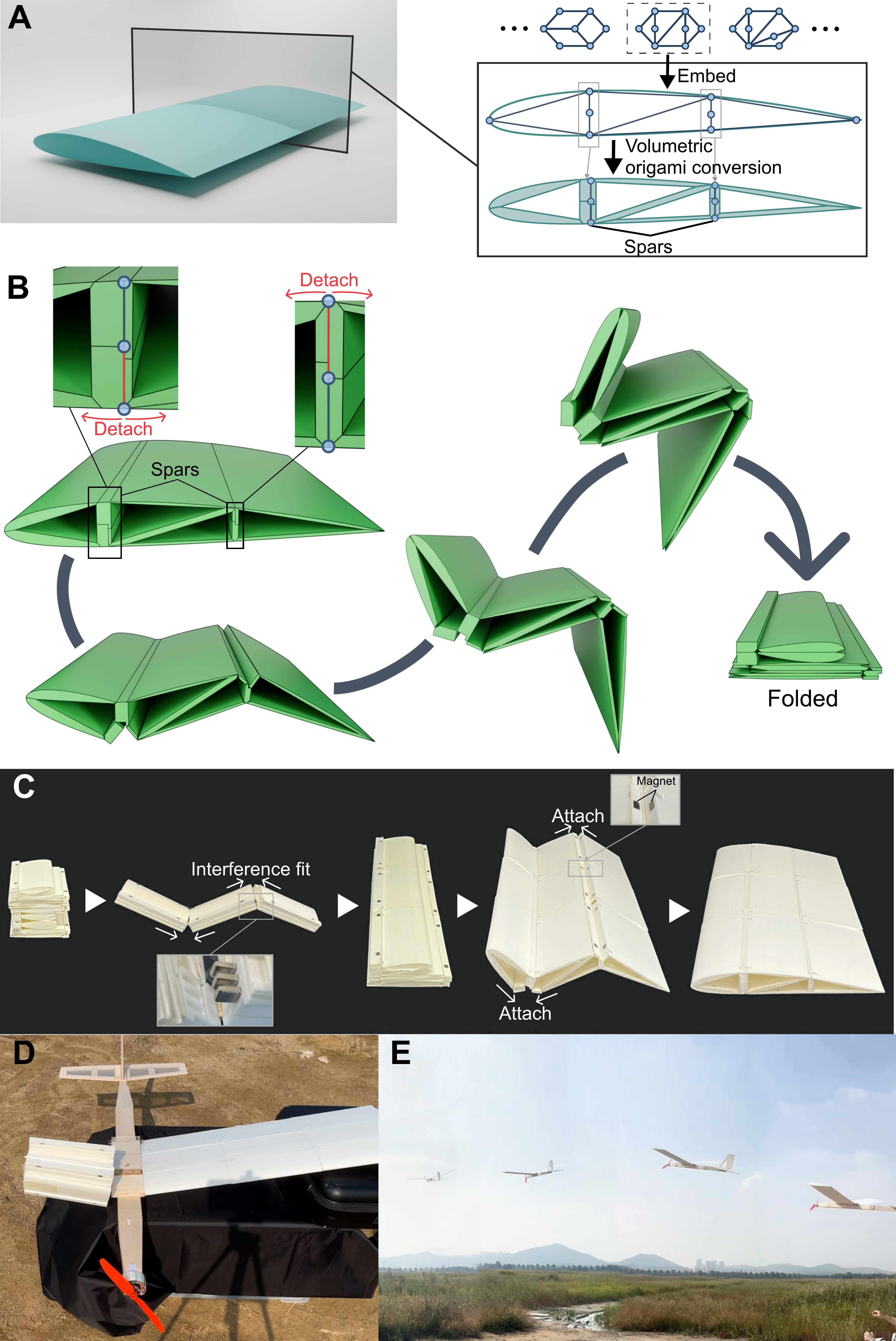}
\caption{
\textbf{Design and flight test of a deployable origami wing.} 
(\textbf{A}) Process of converting an aircraft wing with a NACA 2412 airfoil into a flat-foldable volumetric origami. The origami layout is deliberately selected and embedded to incorporate two vertical spars within the structure.
(\textbf{B}) Flat-folding motion of the volumetric origami wing targeting the NACA 2412 airfoil. Panels detach along the internal edges of the embedded origami layout.
(\textbf{C}) Deployment sequence of the origami wing. The first three stages show the spanwise folding process in reverse, while the final two stages represent the reverse of flat-folding.
(\textbf{D}) Prototype UAV equipped with the deployable origami wing. The left wing is stowed, while the right wing is fully deployed for flight testing.
(\textbf{E}) Flight trajectory of the UAV equipped with the deployable origami wing during flight testing.
}
    \label{d_wing}
\end{figure*}
\restoregeometry


\clearpage

\nocite{McKay1998,McKayPiperno2014,PolyaRead1987}

\bibliography{main}
\bibliographystyle{sciencemag}

\clearpage

\section*{Acknowledgments}

\paragraph*{Funding:}
We acknowledge the support from the Asian Office of Aerospace Research and Development (FA2386-24-1-4051), SNU-IAMD, SNU-IOER, and National Research Foundation grants funded by the Korean government (2023R1A2C2003705 and 2022H1D3A2A03096579).

\paragraph*{Author contributions:}

B.K., Y.M., and J.Y. conceptualized the work; B.K. conceived the inverse design and performed the analysis on the contraction ratio. G.C. performed canonical augmentation to generate valid origami layouts. B.K., H.L., and J.Y. designed and fabricated the origami wing and UAV, and conducted the flight test. All authors extensively contributed to the work and finalizing the manuscript. J.Y. supervised the project.

\paragraph*{Competing interests:}
B.K., H.L., and J.Y. are inventors on a Korean patent application related to this work filed by Seoul National University and Inha University (no. 10-2025-0177686, filed on 21 November 2025). All other authors declare no competing interests.

\paragraph*{Data, code, and materials availability:}
All data and code needed to evaluate and reproduce the results in the paper are present in the paper and/or the Supplementary Materials.
This study did not generate new materials.
Data supporting the findings of this study are available in the Zenodo repository under the accession code 
\href{https://doi.org/10.5281/zenodo.20439881}{https://doi.org/10.5281/zenodo.20439881}.


\subsection*{Supplementary materials}
Supplementary Text\\
Figs. S1 to S15\\
Tables S1 to S6\\
Movie~S1 to S2


\newpage


\renewcommand{\thefigure}{S\arabic{figure}}
\renewcommand{\thetable}{S\arabic{table}}
\renewcommand{\theequation}{S\arabic{equation}}
\renewcommand{\thealgocf}{S\arabic{algocf}}
\renewcommand{\thepage}{S\arabic{page}}
\renewcommand\thesubsection{Supplementary Note \arabic{subsection}}
\renewcommand\thesubsubsection{Supplementary Note \arabic{subsection}.\arabic{subsubsection}}
\setcounter{figure}{0}
\setcounter{table}{0}
\setcounter{equation}{0}
\setcounter{page}{1} 

\titleformat{\part}[display]
  {\normalfont\huge\bfseries\centering} 
  {\partname\ \thepart}                 
  {12pt}                                
  {\Large}                               
\titleformat{\subsection}
  {\normalfont\normalsize\bfseries} 
  {\thesubsection}                 
  {1em}                            
  {}                               

\begin{center}
\part*{Supplementary Materials for\\ \scititle}

	Byoung-Gyu~Kim$^{1}$,
	Geonhee~Cho$^{1}$,
	Yasuhiro~Miyazawa$^{1}$,
    Hak-Tae~Lee$^{2}$,
    Jinkyu~Yang$^{1\ast}$
    
	\small$^{1}$Department of Mechanical Engineering, Seoul National University, 1 Gwanak-ro, Gwanak-gu, Seoul 08826, South Korea
    
	\small$^{2}$Department of Aerospace Engineering and the Program in Aerospace Systems Convergence, Inha University, 100 Inha-ro, Michuhol-gu, Incheon 22212, South Korea
    
	\small$^\ast$Corresponding author. Email: jkyang11@snu.ac.kr
\end{center}

\subsubsection*{This PDF file includes:}
Supplementary Text\\
Figures S1 to S15\\
Tables S1 to S6\\
Captions for Movies S1 to S2

\subsubsection*{Other Supplementary Materials for this manuscript:}
Movies S1 to S2

\newpage


\section*{Supplementary Text}

\subsection{Efficient origami layout enumeration}

Before enumerating origami layouts, we first defined the conditions that characterize a valid origami layout.
We assume that valid origami layouts satisfy the following conditions:
\begin{enumerate}\itemsep2pt
\item[(i)] Planar maximal bipartite,
\item[(ii)] Each internal face contains at least one inner vertex,
\item[(iii)] Each internal face contains at least one outer edge.
\end{enumerate}
Condition (i) ensures that all internal faces are quadrilaterals.
Since each edge corresponds to a panel, each vertex to a crease, and each internal face to a thick origami cell, each cell consists of four panels connected by four creases (i.e., a four-bar linkage–like structure), and thus possesses the necessary degree of freedom.
Additionally, the one degree of freedom enforced by the four-bar linkage-like structure, together with condition (ii), guarantees that the entire thick origami structure can be locked in the deployed state by temporarily bonding the unbonded interface, as shown in the last subsection of Supplementary Note 2.
Condition (iii) was imposed to ensure that all cells participate in representing the curved surface, thereby eliminating redundant internal structures.

\begin{figure*}[ht]
    \centering
    \includegraphics[width=178mm]{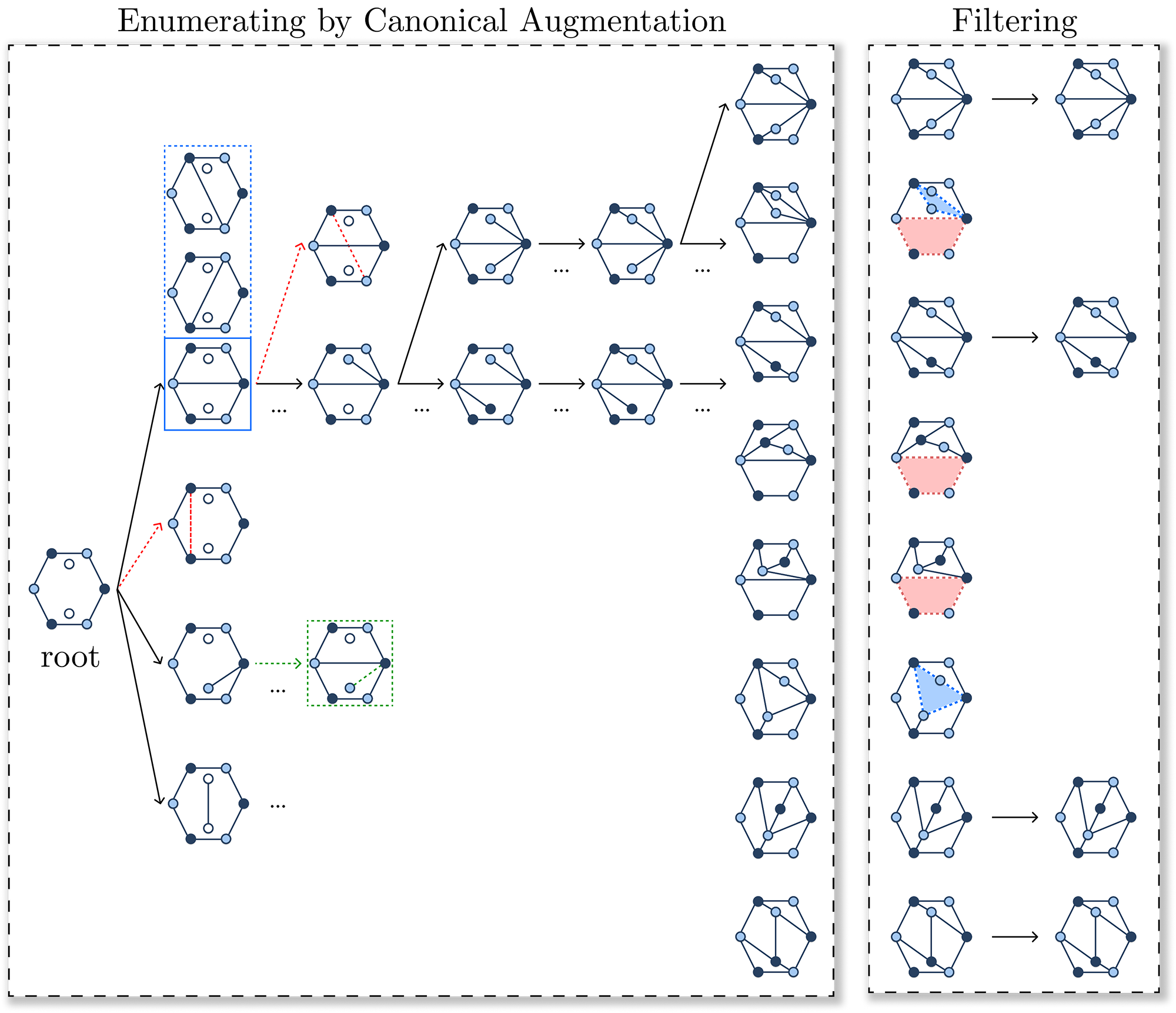}
  \caption{\textbf{Search tree for graph generation under condition}. In the enumeration part, blue dashed boxes indicate the orbit under the symmetry group (graphs equivalent by the action of \(D_{v_{\text{out}}}\times S_{v_{\text{in}}}\)); the blue solid box marks the canonical representative selected by CA. Green dashed boxes denote candidates rejected by the parent test. Red dashed boxes denote candidates pruned during enumeration because they violate the enforced hereditary conditions. 
  In the filtering part, a red‐shaded face indicates a violation of condition (ii), and a blue‐shaded face indicates a violation of condition (iii). Maximality in (i) is automatically satisfied because only CA leaves proceed to filtering.}
    \label{fig:efficiency}
\end{figure*}

To efficiently enumerate all layouts satisfying the condition, while avoiding isomorphic duplication, we adopt canonical augmentation (CA)~\cite{McKay1998}, corresponding to the Enumeration part of Fig.~\ref{fig:efficiency}. CA is a search–tree procedure that generates graphs by adding one admissible edge at a time from a chosen root graph. It prevents isomorphic duplicates by coupling a deletion–based augmentation rule with canonical labeling~\cite{McKay1998}: the deletion rule ensures that each graph is generated only once from a canonical parent, and canonical labeling (computed using nauty/Traces~\cite{McKayPiperno2014}) identifies canonical forms and prunes the search space. As described in~\cite{McKay1998}, to efficiently prune the search space during the augmentation process, one can apply additional conditions, and the applied conditions must be Hereditary Property. A property $\mathcal{P}$ is hereditary if $G \in \mathcal{P}$ implies $G' \in \mathcal{P}$ for all subgraphs $G' \subseteq G$. 

The root graph is the outer cycle \(C_{v_{\text{out}}}\) with \(v_{\text{out}}\) outer vertices together with \(v_{\text{in}}\) isolated inner vertices. This minimal structure ensures the outer panels form a closed shape ($C_{v_{\text{out}}}$) and secures the locking potential of internal cells (placing inner vertices). The largest symmetry that preserves this structure is \(G = D_{v_{\text{out}}} \times S_{v_{\text{in}}}\), where \(D_{v_{\text{out}}}\) acts on \(C_{v_{\text{out}}}\) by rotations and reflections and \(S_{v_{\text{in}}}\) permutes indistinguishable inner vertices. Under the action of \(G\), CA generates exactly one representative per orbit, which substantially prunes the search tree.

During generation, we enforce bipartiteness and planarity, since both are hereditary (i.e., any subgraph of a bipartite graph is bipartite, and any subgraph of a planar graph is planar) and therefore suitable for early pruning~\cite{McKay1998}. By contrast, maximality in (i) and the face–incidence requirements (ii)–(iii) depend on the completed planar embedding: deleting a single edge may merge two quadrilaterals into a larger face and invalidate maximality, and intermediate graphs do not possess well–defined final faces. These properties are applied only once, as a post–generation filter on terminal candidates. This two–stage pipeline is summarized in Fig.~\ref{fig:efficiency}, where the Filtering part corresponds to the process mentioned above.

\begin{figure*}[ht]
    \centering
    \includegraphics[width=178mm]{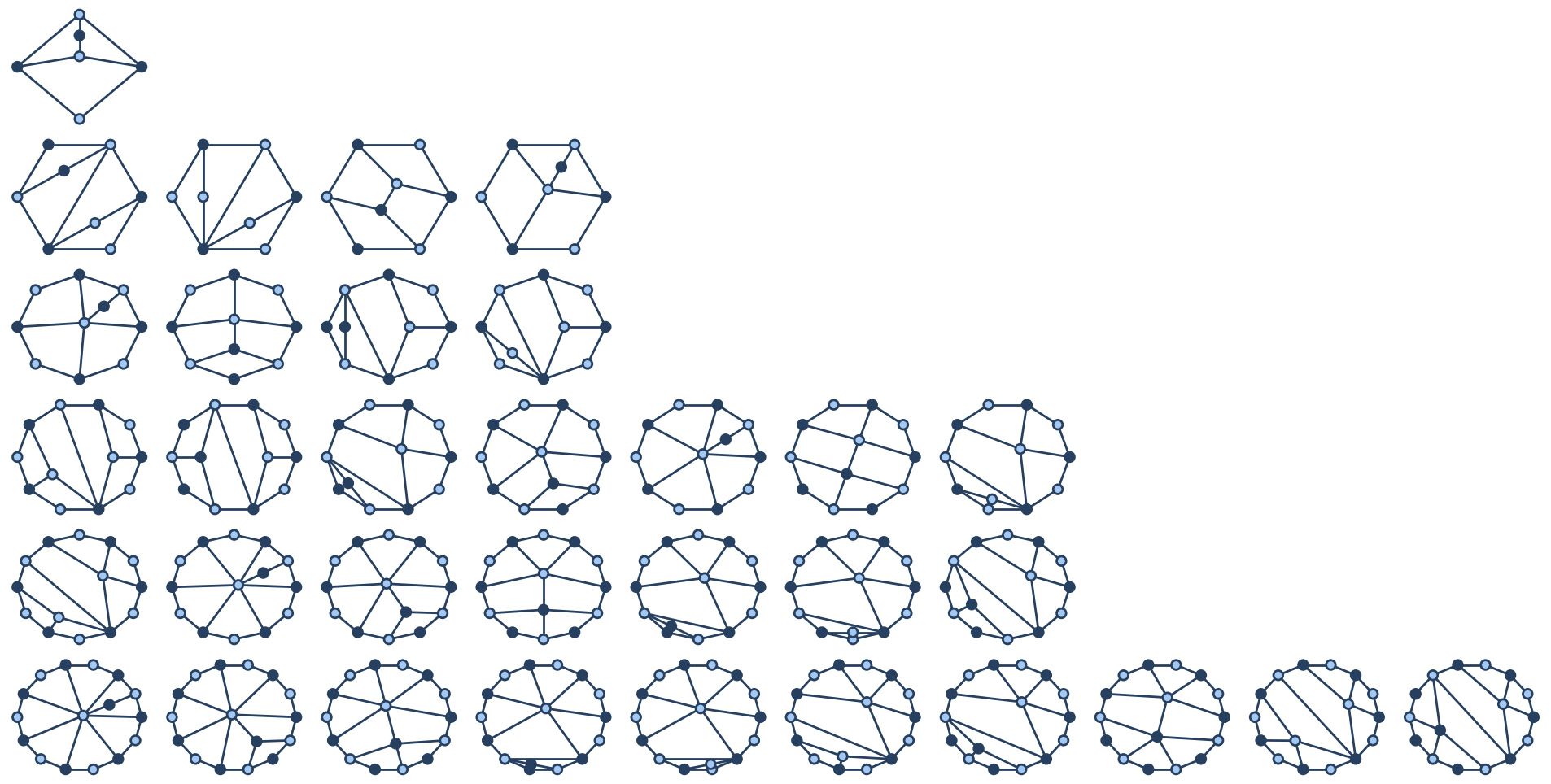}
  \caption{\textbf{Representative non-isomorphic layouts}. Each row corresponds to one boundary size $v_{\mathrm{out}}\in\{4,6,\dots,14\}$ (increasing downward) for fixed $v_{\mathrm{in}}=2$.; within a row, distinct isomorphism classes satisfying conditions (i)–(iii) are shown.}
    \label{fig:gallery}
\end{figure*}

\begin{figure*}[ht]
    \centering
    \includegraphics[width=178mm]{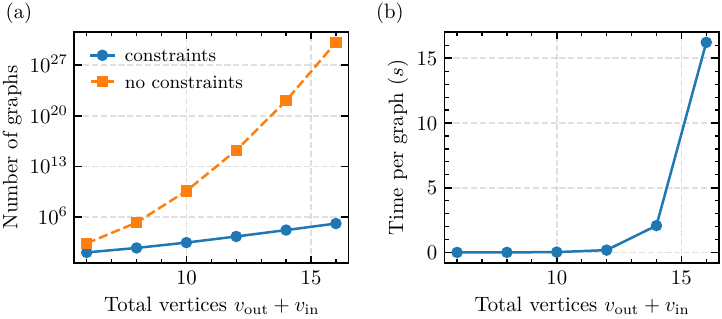}
  \caption{(\textbf{A}) \textbf{Search–space compression from enforcing constraints during CA.} Plotted is the number of non-isomorphic graphs explored versus total vertices $v_{\mathrm{out}}+v_{\mathrm{in}}$ (log scale). Constrained CA (blue) indicates the search space of conditioned CA along total vertices. Unconstrained CA (orange) indicates the search space of unconditioned CA along total vertices. 
 (\textbf{B}) \textbf{Computational cost per valid layout.} Total runtime divided by the target count.}
    \label{fig:comparing}
\end{figure*}

This pipeline yields the final set of non–isomorphic layouts; a representative gallery appears in Fig.~\ref{fig:gallery}. Fig.~\ref{fig:comparing}A shows the reduction of the explored search space relative to unconstrained CA, using a symmetry–aware baseline count computed via the Pólya enumeration theorem (PET)~\cite{PolyaRead1987}; Fig.~\ref{fig:comparing}B reports the generating time of each final layout, indicating that time per valid graph remains feasible. All timings in Fig.~\ref{fig:comparing}B were obtained on a single commodity CPU–only workstation (x86\_64 Linux under WSL2; kernel~6.6.87, glibc~2.39) with an Intel Core i7–13700F (12~cores / 24~threads, AVX2) and 16~GB RAM; no discrete GPU. The implementation uses Python~3.12.11 linked against OpenBLAS~0.3.29 (64–bit integers, dynamic–arch “Haswell”), NetworkX~3.5, and pynauty~2.8.8.1. Thread caps (OMP/MKL/OPENBLAS) were left at defaults.

\subsection{Inverse design for generalized cylinders}
\subsubsection*{Thickness-dependent flat-foldability constraints}
In this section, we derive explicit expressions for $T_1(s_2')$, $T_2(s_2')$, $T_3(s_4')$, and $T_4(s_4')$ in Fig.~\ref{fig:cylinder_thickness}A.
We assume that the coordinates of the four vertices $\mathbf{q}_1$, $\mathbf{q}_2$, $\mathbf{q}_3$, and $\mathbf{q}_4$, as well as the target cross-section $\mathbf{f}(s)$, are given, and that all position vectors lie in the $xy$-plane.
Note that $\mathbf{q}_1$ is always an internal vertex in the origami layout, and the indices of $\mathbf{q}_i$ are assigned in a counterclockwise order.

To derive $T_1(s_2')$, we first define the distance $d_1$, as shown in Fig.~\ref{fig:cylinder_thickness}B, which is the shortest distance between $\mathbf{q}_2'$ and the line passing through $\mathbf{q}_1$ and $\mathbf{q}_2$.
Accordingly, $d_1$ is given by $d_1 = \lVert \mathbf{q}_2' - \mathbf{q}_1 \rVert \sin \psi_1$.
Since $\lVert (\mathbf{q}_2-\mathbf{q}_1) \times (\mathbf{q}_2'-\mathbf{q}_1) \rVert
= \lVert \mathbf{q}_2-\mathbf{q}_1 \rVert \, \lVert \mathbf{q}_2'-\mathbf{q}_1 \rVert \sin \psi_1$, $d_1$ is given as follows:
\begin{equation}
\label{eq:d1_1}
d_1 = \frac{\lVert (\mathbf{q}_2-\mathbf{q}_1) \times (\mathbf{q}_2'-\mathbf{q}_1) \rVert}{\lVert \mathbf{q}_2-\mathbf{q}_1 \rVert}.
\end{equation}
Since all position vectors lie in the $xy$-plane,  $(\mathbf{q}_2-\mathbf{q}_1) \times (\mathbf{q}_2'-\mathbf{q}_1)
= \det[\mathbf{q}_2-\mathbf{q}_1 \;\; \mathbf{q}_2'-\mathbf{q}_1]\,\mathbf{k}$ ($\mathbf{k}$ is the unit vector along the $z$-axis).
Therefore, $d_1$ is given by
\begin{equation}
\label{eq:d1_2}
d_1 = \frac{\det\!\big[\mathbf{q}_2-\mathbf{q}_1 \;\; \mathbf{q}_2'-\mathbf{q}_1\big]}{\lVert \mathbf{q}_2-\mathbf{q}_1 \rVert}.
\end{equation}
Here, since $\mathbf{q}_1$, $\mathbf{q}_2$, $\mathbf{q}_3$, and $\mathbf{q}_4$ are assigned in a counterclockwise order and $\mathbf{q}_2'$ always lies between $\mathbf{q}_2$ and $\mathbf{q}_3$, $\mathbf{q}_2'$ is always located on the left-hand side of the vector $\mathbf{q}_2 - \mathbf{q}_1$, as shown in Fig.~\ref{fig:cylinder_thickness}A.
Therefore, the determinant $\det[\mathbf{q}_2-\mathbf{q}_1 \;\; \mathbf{q}_2'-\mathbf{q}_1]$ is always positive, and thus the absolute value is omitted.
We assume that the orange dashed line in Fig.~\ref{fig:cylinder_thickness}B is always parallel to $\mathbf{q}_2-\mathbf{q}_1$.
Therefore, the panel thickness $T_1(s_2')$ satisfies $T_1(s_2') = d_1$, and since $\mathbf{q}_2' = \mathbf{f}(s_2')$, $T_1(s_2')$ can be written as follows:
\begin{equation}
\label{eq:T1}
T_1(s_2') = \frac{\det\!\big[\mathbf{q}_2-\mathbf{q}_1 \;\; \mathbf{f}(s_2')-\mathbf{q}_1\big]}{\lVert \mathbf{q}_2-\mathbf{q}_1 \rVert}.
\end{equation}

To derive an explicit expression for $T_2(s_2')$, as shown in Fig.~\ref{fig:cylinder_thickness}C, we first define $d_2(s)$ as the distance between a point of $\mathbf{f}(s)$ and the line passing through $\mathbf{q}_2'$ and $\mathbf{q}_3$.
Following the same procedure as for $d_1$, $d_2(s)$ can be expressed as follows:
\begin{equation}
\label{eq:d2(s)}
d_2(s) = \frac{\det\!\big[\mathbf{q}_2'-\mathbf{q}_3 \;\; \mathbf{f}(s)-\mathbf{q}_3\big]}{\lVert \mathbf{q}_2'-\mathbf{q}_3 \rVert}.
\end{equation}
Since the panel thickness $T_2(s_2')$ in Fig.~\ref{fig:cylinder_thickness}C is given by the maximum value of $d_2(s)$, it can be expressed as follows:
\begin{equation}
\label{eq:T2}
T_2(s_2') = \max_{\,s_3 \le s \le s_2'} 
\frac{\det\!\big[\mathbf{f}(s_2')-\mathbf{q}_3 \;\; \mathbf{f}(s)-\mathbf{q}_3\big]}
{\lVert \mathbf{f}(s_2')-\mathbf{q}_3 \rVert}.
\end{equation}
Note that $\mathbf{q}_2' = \mathbf{f}(s_2')$, $\mathbf{q}_3 = \mathbf{f}(s_3)$, and that we assume $s_3 \le s_2'$.
The expression for $d_2(s)$ in Eq.~\eqref{eq:d2(s)} provides sign information indicating whether $\mathbf{f}(s)$ lies inside or outside the line connecting $\mathbf{q}_2'$ and $\mathbf{q}_3$, as illustrated in Fig.~\ref{fig:cylinder_thickness}D.
This sign information ensures that Eq.~\eqref{eq:T2} always selects the thickness $T_2(s_2')$ corresponding to the outward protrusion distance of $\mathbf{f}(s)$ measured with respect to the line connecting $\mathbf{q}_2'$ and $\mathbf{q}_3$.
As shown in Fig.~\ref{fig:cylinder_thickness}E, this choice of $T_2(s_2')$ ensures that $T_1(s_2')$ and $T_2(s_2')$ measure thicknesses on the same side.
This ensures that the thickness-dependent flat-foldability constraint given in Eq.~(1) of the main text guarantees a constant thickness through the selection of appropriate thicknesses.
Similarly, $T_3(s_4')$ and $T_4(s_4')$ can be obtained following the same procedure used to derive $T_1(s_2')$ and $T_2(s_2')$, and are given as follows:

\begin{equation}
\label{eq:T3}
T_3(s_4') = \max_{\,s_4' \le s \le s_3}
\frac{\det\!\big[\mathbf{f}(s)-\mathbf{q}_3 \;\; \mathbf{f}(s_4') - \mathbf{q}_3\big]}
{\lVert \mathbf{f}(s_4') - \mathbf{q}_3 \rVert},
\end{equation}

\begin{equation}
\label{eq:T4}
T_4(s_4') =\frac{\det\!\big[\mathbf{f}(s_4') - \mathbf{q}_1 \;\; \mathbf{q}_4 - \mathbf{q}_1\big]}
{\lVert \mathbf{q}_4 - \mathbf{q}_1 \rVert}.
\end{equation}
Using the explicit expressions for $T_1(s_2')$, $T_2(s_2')$, $T_3(s_4')$, and $T_4(s_4')$, we numerically solve the thickness-dependent flat-foldability constraints in the main text (Eqs.~(1) and~(2)) using Brent's method implemented in \texttt{scipy.optimize.brentq}.

In the above procedure, we assume that $\mathbf{q}_2$, $\mathbf{q}_3$, and $\mathbf{q}_4$ all lie on $\mathbf{f}(s)$, i.e., the edges $\mathbf{q}_2\mathbf{q}_3$ and $\mathbf{q}_3\mathbf{q}_4$ are outer edges, while $\mathbf{q}_1$ is always an internal vertex.
However, as shown in Fig.~\ref{fig:cylinder_thickness}F, either of the edges $\mathbf{q}_2\mathbf{q}_3$ or $\mathbf{q}_3\mathbf{q}_4$ may instead be an internal edge.
In this case, we simply set $\mathbf{q}_2' = \mathbf{q}_2$ or $\mathbf{q}_4' = \mathbf{q}_4$, depending on which side corresponds to the internal edge, since no thickness is defined for an internal edge by $\mathbf{f}(s)$.

\subsubsection*{Lengthwise flat-foldability constraint}
In this section, we derive the explicit form of Eq.~(3) in the main text.
We begin by defining the admissible ranges of the offset distances $t_1$ and $t_2$.
When $\mathbf{f}(s)$ is convex, the offset distances $t_1$ and $t_2$ are non-negative, which ensures that the offset is directed inward.
However, as shown in Fig.~\ref{fig:cylinder_length}A, when $\mathbf{f}(s)$ contains a concave region, the offset distance (in this case, $t_1$) must exceed the maximum inward indentation, ensuring that the panel remains a single connected piece.
As shown in Fig.~\ref{fig:cylinder_thickness}D, inward indentation occurs when $d_2(s) < 0$.
Therefore, the maximum inward indentation is given by $\left|\min_{s_3 \le s \le s'_2} d_2(s)\right|$. 
Using Eq.~\eqref{eq:d2(s)}, the admissible range of $t_1$ is then given as follows:

\begin{equation}
\label{eq:t1_scope}
t_1 \ge -\min_{s_3 \le s \le s'_2} \frac{\det\!\big[\mathbf{q}_2'-\mathbf{q}_3 \;\; \mathbf{f}(s)-\mathbf{q}_3\big]}{\lVert \mathbf{q}_2'-\mathbf{q}_3 \rVert}.
\end{equation}
Similarly, $t_2$ must satisfy the following constraint:
\begin{equation}
\label{eq:t2_scope}
t_2 \ge -\min_{s'_4 \le s \le s_3} \frac{\det\!\big[\mathbf{f}(s)-\mathbf{q}_3 \;\; \mathbf{q}_4'  - \mathbf{q}_3\big]}
{\lVert \mathbf{q}_4' - \mathbf{q}_3 \rVert}.
\end{equation}
Note that Eqs.~\eqref{eq:t1_scope} and~\eqref{eq:t2_scope} reduce to $t_1 \ge 0$ and $t_2 \ge 0$, respectively, when there is no inward indentation.

To derive explicit expressions for $l_1(t_1,t_2)$ and $l_2(t_1,t_2)$, we first define $\delta_1$, $\delta_2$, $\delta_3$, and $\delta_4$ as shown in Fig.~\ref{fig:cylinder_length}B.
Thus, we can express $l_1(t_1,t_2)$ and $l_2(t_1,t_2)$ as follows:
\begin{equation}
\label{eq:l1_first}
l_1(t_1,t_2)
=
\left\| \mathbf{q}_1' - \mathbf{q}_2' \right\|
+
\left\| \mathbf{q}_3 - \mathbf{q}_2' \right\|
-
\delta_1
-
\delta_2,
\end{equation}
\begin{equation}
\label{eq:l2_first}
l_2(t_1,t_2)
=
\left\| \mathbf{q}_1' - \mathbf{q}_4' \right\|
+
\left\| \mathbf{q}_3 - \mathbf{q}_4' \right\|
-
\delta_3
-
\delta_4.
\end{equation}
Note that $\delta_i$ denotes a signed distance.
For example, in Fig.~\ref{fig:cylinder_length}B, $\delta_1$, $\delta_3$, and $\delta_4$ are positive, whereas $\delta_2$ is negative, so that $l_1(t_1,t_2)$ and $l_2(t_1,t_2)$ are correctly evaluated.
Furthermore, from the geometry shown in Fig.~\ref{fig:cylinder_length}C, we can express $\delta_1$ as follows:
\begin{equation}
\label{eq:delta1}
\delta_1
=
\frac{t_1}{\tan\alpha}+\frac{t_2}{\sin\alpha}.
\end{equation}
Note that $\mathbf{q}_1'$ denotes the position of $\mathbf{v}_1$ before the offset, i.e., the position of $\mathbf{v}_1$ at Stage~III in Fig.~2A of the main text.
Similarly, the remaining $\delta_i$ can be expressed as follows:
\begin{equation}
\label{eq:delta2}
\delta_2
=
\frac{t_1}{\tan\beta}+\frac{t_2}{\sin\beta},
\end{equation}
\begin{equation}
\label{eq:delta3}
\delta_3
=
\frac{t_2}{\tan\beta}+\frac{t_1}{\sin\beta},
\end{equation}
\begin{equation}
\label{eq:delta4}
\delta_4
=
\frac{t_2}{\tan\alpha}+\frac{t_1}{\sin\alpha}.
\end{equation}
By substituting Eqs.~\eqref{eq:delta1}--\eqref{eq:delta4} into Eqs.~\eqref{eq:l1_first} and \eqref{eq:l2_first}, we obtain the lengthwise flat-foldability constraint (Eq.~(3) in the main text), which can be written as follows:
\begin{equation}
\label{eq:cylinder_length}
t_2 - t_1
=
\frac{\left\| \mathbf{q}_1' - \mathbf{q}_4' \right\|
+
\left\| \mathbf{q}_3 - \mathbf{q}_4' \right\|
-
\left\| \mathbf{q}_1' - \mathbf{q}_2' \right\|
-
\left\| \mathbf{q}_3 - \mathbf{q}_2' \right\|}{\cot\alpha+\cot\beta-\csc\alpha-\csc\beta}
.
\end{equation}
Note that $\cot\alpha = \frac{1}{\tan\alpha}$ and $\csc\alpha = \frac{1}{\sin\alpha}$.
By choosing $t_1$ and $t_2$ to satisfy inequalities~\eqref{eq:t1_scope} and~\eqref{eq:t2_scope} as well as Eq.~\eqref{eq:cylinder_length}, we can ensure flat-foldability.

However, since $\mathbf{q}_1'$ is defined as the intersection of the inner faces of the panel (indicated by dashed lines) as shown in Fig.~\ref{fig:cylinder_length}A, it is not well-defined when $\mathbf{q}_1$, $\mathbf{q}_2$, and $\mathbf{q}_4$ are collinear, as in Fig.~\ref{fig:cylinder_length}D, because the dashed lines are parallel and thus no intersection exists.
Therefore, when $\mathbf{q}_1$, $\mathbf{q}_2$, and $\mathbf{q}_4$ are collinear, Eq.~\eqref{eq:cylinder_length} cannot be directly applied, as it requires $\mathbf{q}_1'$.
In this case, we instead choose $t_1$ and $t_2$ such that the resulting inward faces of the two panels are collinear (see the second stage of Fig.~\ref{fig:cylinder_length}D), allowing $\mathbf{v}_1$ to be placed.
However, in this case, as shown in the final stage of Fig.~\ref{fig:cylinder_length}D, $\mathbf{v}_1$ cannot be uniquely determined along the line.
Thus, we use Eq.~(3) in the main text to determine the location of $\mathbf{v}_1$ by numerically solving $l_1(\mathbf{v}_1) = l_2(\mathbf{v}_1)$.

\subsubsection*{Locking in the deployed state}
All flat-foldable thick origami cells consist of thick panels, and due to contact between these thick panels, the folding angle is necessarily restricted to a finite range.
For example, as shown in Fig.~\ref{fig:locking}A, when the folding angle $\rho_i$ at $\mathbf{v}_1$ of the $i$-th cell is defined as the dihedral angle between the two planes meeting at $\mathbf{v}_1$, $\rho_i$ is bounded within the following range
\begin{equation}
\label{eq:folding_angle_range}
0\le \rho_i \le \rho_{i,max}.
\end{equation}
In addition, as shown in the first figure of Fig.~\ref{fig:locking}A, due to the parallel relationship induced by the offset, $\rho_{i,\max}$ is equal to $\angle \mathbf{q}_4 \mathbf{q}_1 \mathbf{q}_2$.
According to the definition given in the subsection ``Thickness-dependent flat-foldability constraints,'' $\mathbf{q}_1$ is always an internal vertex of the origami layout.
Therefore, as shown in the first figure of Fig.~\ref{fig:locking}B,
\begin{equation}
\label{eq:max_rho_relationship}
\sum_{i} \rho_{i,\max} = 2\pi,
\end{equation}
where the summation is taken over all cells sharing the same internal vertex.
As shown in the bottom of Fig.~\ref{fig:locking}B, if all interfaces between adjacent cells are bonded to form a thick origami structure, then, since each bonding interface corresponds to an internal edge, $\sum_{i} \rho_i = 2\pi$ must be satisfied.
Thus, from Eqs.~\eqref{eq:folding_angle_range}, \eqref{eq:max_rho_relationship} and $\sum_{i} \rho_i = 2\pi$, it follows that $\rho_i = \rho_{i,\max}$ is enforced.
This implies that all cells are locked in the deployed state, because, as assumed in Supplementary Note 1, each cell behaves as a four-bar linkage and has one degree of freedom.
Accordingly, as shown in the top of Fig.~\ref{fig:locking}B, we intentionally left one internal edge unbonded, which ensures the necessary degree of freedom to fold the thick origami structure.
On the other hand, when we desire to lock the thick origami structure in its deployed state, we temporarily bond the unbonded interfaces, as shown in the deployed wing in the main text.

\subsection{Flat-foldability constraints in spherical geometry}
In this Supplementary Note, we explain in detail how the flat-foldability constraints defined for a generalized cylinder are transformed into a spherical geometry form for a generalized cone.
The main difference between Euclidean geometry (geometry on a plane) and spherical geometry is that parallel lines do not exist in spherical geometry.
Therefore, unlike a generalized cylinder, in which internal facets can be determined from origami layout vertices and crease positions by parallelism, in a generalized cone using spherical geometry, each internal face must be specified.
For example, when $\mathbf{g}(s_4')$, a candidate for $\mathbf{p}_4'$, is given, we define each internal face as shown in Fig.~\ref{fig:cone_thickness}A, where $\frac{\mathbf{p}_4 \cdot \hat{\mathbf{z}}}{\mathbf{p}_1 \cdot \hat{\mathbf{z}}}\,\mathbf{p}_1 - \mathbf{p}_4$ is illustrated in Fig.~\ref{fig:cone_thickness}B.
Each vector illustrated in Fig.~\ref{fig:cone_thickness}A represents the normal vector direction of the internal facet.
The reason for using $\frac{\mathbf{p}_4 \cdot \hat{\mathbf{z}}}{\mathbf{p}_1 \cdot \hat{\mathbf{z}}}\,\mathbf{p}_1 - \mathbf{p}_4$ in defining the normal vector of the internal facet is to ensure that, the intersection of the internal facet with the $xy$-plane is parallel to the intersection between the $xy$-plane and the facet passing through $\mathbf{p}_1$ and $\mathbf{p}_4$.
Similarly, the normal vectors of the two internal facets meeting at $\mathbf{g}(s_2')$ are determined as $\left(\mathbf{p}_2 - \frac{\mathbf{p}_2 \cdot \hat{\mathbf{z}}}{\mathbf{p}_1 \cdot \hat{\mathbf{z}}}\mathbf{p}_1\right) \times \mathbf{g}(s_2')$ and $ \mathbf{p}_3 \times \mathbf{g}(s_2')$.

The thickness-dependent flat-foldability constraints in Eqs.~(4) and (5) in the main text correspond to the condition illustrated in Fig.~\ref{fig:cone_thickness}C. 
In this condition, when the folding angle at $\mathbf{g}(s_4')$ reaches $180^\circ$, the panels are bounded by the plane defined by $\mathbf{p}_1$ and $\mathbf{p}_4$, whereas when the folding angle at $\mathbf{g}(s_2')$ reaches $180^\circ$, the panels are bounded by the plane defined by $\mathbf{p}_1$ and $\mathbf{p}_2$.
Here, $\mathbf{R}(\mathbf{a}, b)$ denotes the rotation matrix given by Rodrigues' formula, corresponding to a rotation about the axis $\mathbf{a}$ by an angle $b$.
In Eqs.~(4) and (5) in the main text, the rotation angles $\phi_{\mathbf{g}(s_2')}$ and $\phi_{\mathbf{g}(s_4')}$ are the dihedral angles between the internal facets meeting at $\mathbf{g}(s_2')$ and $\mathbf{g}(s_4')$, respectively (for example, for $\phi_{\mathbf{g}(s_4')}$, see Fig.~\ref{fig:cone_thickness}A).
Note that the dihedral angles are defined by the angle between the normal vectors of the internal facets.
Thus, $\mathbf{R}(-\mathbf{g}(s_4'), \phi_{\mathbf{g}(s_4')})\,\mathbf{g}(s)$ in Eq.~(5) in the main text represents the position of $\mathbf{g}(s)$ when the folding angle at $\mathbf{g}(s_4')$ is $180^\circ$ (i.e., internal facets (i) and (ii) in Fig.~\ref{fig:cone_thickness}A are coplanar).
Furthermore, $\mathbf{R}(-\mathbf{g}(s_4'), \phi_{\mathbf{g}(s_4')})\,\mathbf{g}(s) \cdot (\mathbf{p}_1 \times \mathbf{p}_4)$ represents a signed measure of how far $\mathbf{g}(s)$ from the plane defined by $\mathbf{p}_1$ and $\mathbf{p}_4$ in the flat-folded state. 
Therefore, Eq.~(5) in the main text corresponds to the condition that the plane defined by $\mathbf{p}_1$ and $\mathbf{p}_4$ is tangent to the panel corresponding to internal facet (ii) in Fig.~\ref{fig:cone_thickness}A in the flat-folded state shown in Fig.~\ref{fig:cone_thickness}C.

As in the inverse design case for a generalized cylinder, thickness-dependent flat-foldability constraints alone do not guarantee flat-folding. 
Therefore, as in (III) of Fig.~3B in the main text, the internal facets are adjusted to satisfy the lengthwise flat-foldability constraint.
In this case, since spherical geometry does not admit translation, translation as used in the generalized cylinder cannot be applied.
Instead, each internal facet is adjusted by rotating it about an axis passing through the apex $\mathbf{o}$.
Fig.~\ref{fig:cone_length}A illustrates the rotation axis for each panel.
Here, the choice of the rotation axes $\frac{\mathbf{p}_4 \cdot \hat{\mathbf{z}}}{\mathbf{p}_1 \cdot \hat{\mathbf{z}}}\,\mathbf{p}_1 - \mathbf{p}_4$ and $\mathbf{p}_2 - \frac{\mathbf{p}_2 \cdot \hat{\mathbf{z}}}{\mathbf{p}_1 \cdot \hat{\mathbf{z}}}\mathbf{p}_1$ preserves the property that the intersection of each internal facet with the $xy$-plane is parallel to the intersection between the $xy$-plane and the facet passing through $\mathbf{p}_1$ and $\mathbf{p}_4$, or to that between the $xy$-plane and the facet passing through $\mathbf{p}_2$ and $\mathbf{p}_1$.
Also, the choice of the remaining two rotation axes preserves the thickness-dependent flat-foldability constraints by ensuring that, even after adjusting the internal facets, rotating by $\phi_{\mathbf{p}_2'}$ about crease $\mathbf{p}_2'$ and by $\phi_{\mathbf{p}_4'}$ about crease $\mathbf{p}_4'$ renders the adjacent internal facets coplanar (see Fig.~\ref{fig:cone_length}B).
In this flat-folded state of the two panels adjacent to crease $\mathbf{p}_4'$ shown in Fig.~\ref{fig:cone_length}B, the spherical distance between creases $\mathbf{u}_1$ and $\mathbf{u}_3$ is given by $\cos^{-1}\!\left( \mathbf{R}(-\mathbf{p}_4', \phi_{\mathbf{p}_4'})\,\mathbf{u}_3 \cdot \mathbf{u}_1 \right)$.
Similarly, for the two panels adjacent to crease $\mathbf{p}_2'$, the spherical distance is given by $\cos^{-1}\!\left( \mathbf{R}(\mathbf{p}_2', \phi_{\mathbf{p}_2'})\,\mathbf{u}_3 \cdot \mathbf{u}_1 \right)$. For flat-foldability, these two spherical distances must be equal, from which Eq.~(6) in the main text is derived.

\subsection{Analysis of packaging efficiency}
\subsubsection{Numerical optimization of the geometric realization of an origami layout on a cross-section}
In this section, we present the numerical optimization methods used to optimize the geometric realization of an origami layout on a cross-section, aiming to achieve the minimum contraction ratio.
Given an origami layout and the target cross-section, its geometric realization on a cross-section is determined by the positions of all vertices in the layout.
Since the contraction ratio of the resulting flat-foldable volumetric origami is determined by the geometric realization of the origami layout, it can be expressed as follows:
\begin{equation}
\label{eq:contrac_ratio_func}
r_c = r_c(\mathbf{q}_0^{(1)}, \cdots, \mathbf{q}_0^{(a)}, \mathbf{q}_1^*, \cdots, \mathbf{q}_b^*, \mathbf{q}_1, \cdots, \mathbf{q}_c).
\end{equation}
Note that we assume that the additional thicknesses $t_1$ and $t_2$ are minimized.
Here, $\mathbf{q}_0$ denotes the positions of the internal vertices of the origami layout; $\mathbf{q}_i$ denotes the positions of the outer vertices that are connected to internal vertices; and $\mathbf{q}_i^*$ denotes the positions of the outer vertices that are not connected to internal vertices (see Fig.~\ref{fig:analysis_geometry}A).
Since the outer vertices lie on the boundary of the target cross-section $\mathbf{f}(s)$, they can be parameterized as $\mathbf{q}_i = \mathbf{f}(s_i)$ and $\mathbf{q}_i^* = \mathbf{f}(s_i^*)$. Accordingly, the numerical optimization is performed with respect to the following parameters.
\begin{equation}
\label{eq:contrac_ratio_numerical_opt}
\min_{\mathbf{q}_0^{(1)}, \dots, \mathbf{q}_0^{(a)},\, s_1^*, \dots, s_b^*,\, s_1, \dots, s_c}
\; r_c(\mathbf{q}_0^{(1)}, \dots, \mathbf{q}_0^{(a)}, s_1^*, \dots, s_b^*, s_1, \dots, s_c).
\end{equation}
The resulting numerical optimization problem is solved using the SLSQP algorithm implemented in \texttt{scipy.optimize.minimize}.
For the numerical results in Fig. 4 of the main text and Fig.~\ref{fig:analysis_limit}, we fix the internal vertex at the origin of the superellipse.

\subsubsection{Derivation of the optimal contraction ratio in the continuum limit}
In this section, we derive the relationship between the optimal contraction ratio and the number of cells.
We assume that the target shape is a generalized cylinder with a smooth, convex cross-section, and that the origami layout contains a single internal vertex.
We further assume the continuum limit, in which the central angles ($\Delta \theta_i$, see Fig.~\ref{fig:analysis_geometry}A) of individual cells are infinitesimal.
To obtain the optimal contraction ratio, the cross-sectional area of each cell, $a_i$, must be minimized, which in turn requires minimizing the thickness of the panels.
Therefore, to eliminate the unnecessary thicknesses $t_1$ and $t_2$, we assume that, for each cell, the placement of the vertex $\mathbf{q}_i^{*}$ in the origami layout satisfies the lengthwise flat-foldability constraint without the additional thicknesses $t_1$ and $t_2$, together with $\mathbf{q}_0$, $\mathbf{q}_i$, and $\mathbf{q}_{i+1}$.
By estimating $\varepsilon$ in Fig.~\ref{fig:analysis_geometry}A by imposing an approximated form of the lengthwise flat-foldability constraint, $\overline{\mathbf{q}_0 \mathbf{q}_i} + \overline{\mathbf{q}_i \mathbf{q}_i^*} = \overline{\mathbf{q}_0 \mathbf{q}_{i+1}} + \overline{\mathbf{q}_i^* \mathbf{q}_{i+1}}$, we determine the position of $\mathbf{q}_i^{*}$.
Since $\Delta \theta_i$ is very small, the distances $\overline{\mathbf{q}_i \mathbf{q}_i^*}$ and $\overline{\mathbf{q}_i \mathbf{q}_{i+1}^*}$ can be approximated as infinitesimal curve lengths $\overline{\mathbf{q}_i \mathbf{q}_i^*} \simeq \sqrt{r(\theta_i)^2 + r'(\theta_i)^2}\,\varepsilon \Delta \theta_i$ and $\overline{\mathbf{q}_i \mathbf{q}_{i+1}^*} \simeq \sqrt{r(\theta_i)^2 + r'(\theta_i)^2}\,(1-\varepsilon) \Delta \theta_i$, respectively.
Therefore, $\varepsilon$ is determined as follows:
\begin{equation}
\label{eq:epsilon}
\varepsilon = \frac{1}{2} \left( 1 + \frac{r'(\theta_i)}{\sqrt{r(\theta_i)^2 + r'(\theta_i)^2}} \right).
\end{equation}
Next, we approximately impose the thickness-dependent flat-foldability constraints, as shown on the left of Fig.~\ref{fig:analysis_geometry}B, to assign the necessary thickness to each panel.
To determine the thicknesses $T_1$ and $T_3$, we approximate each infinitesimal curve as a circular arc, as shown on the right of Fig.~\ref{fig:analysis_geometry}B.
This gives $\delta = \left|\kappa(\theta_i)\right| \sqrt{r(\theta_i)^2 + r'(\theta_i)^2}\,\varepsilon \Delta \theta_i$. If we regard $T_1$ as a sagitta, the following result is obtained:
\begin{equation}
\label{eq:T1_analysis}
T_1 \simeq \frac{1}{8}\,\frac{1}{\left|\kappa(\theta_i)\right|}\,\delta^2
= \frac{1}{8}\left(r(\theta_i)^2 + r'(\theta_i)^2\right)\left|\kappa(\theta_i)\right| \varepsilon^2 \Delta \theta_i^2.
\end{equation}
Similarly, the thickness $T_3$ can be expressed as follows:
\begin{equation}
\label{eq:T3_analysis}
T_3 \simeq \frac{1}{8}\left(r(\theta_i)^2 + r'(\theta_i)^2\right)\left|\kappa(\theta_i)\right| (1-\varepsilon)^2 \Delta \theta_i^2.
\end{equation}
As $\Delta \theta_i$ becomes small, the width and thickness of panels (ii) and (iii) in Fig.~\ref{fig:analysis_geometry}B diminish, whereas only the thickness of panels (i) and (iv) decreases. Therefore, the cross-sectional areas of panels (ii) and (iii) become higher-order terms and can be neglected. Consequently, the bounded cross-sectional area of each cell, $a_i$, can be approximated as the sum of the areas of panels (i) and (iv), $a_i \simeq T_1 r(\theta_i) + T_3 r(\theta_i + \Delta \theta_i)$.
By substituting the results of Eqs.~\eqref{eq:T3_analysis}, \eqref{eq:T1_analysis}, and \eqref{eq:epsilon}, $a_i$ can be expressed as follows:
\begin{equation}
\label{eq:a_i}
a_i \simeq \frac{1}{16}\, r(\theta_i)\left|\kappa(\theta_i)\right|\left(r(\theta_i)^2 + 2\,r'(\theta_i)^2\right)\Delta \theta_i^2.
\end{equation}
Therefore, the contraction ratio is accordingly given by
\begin{equation}
\label{eq:r_c}
r_c \simeq \frac{\sum_{i=1}^{n} \frac{1}{16}\, r(\theta_i)\left|\kappa(\theta_i)\right|\left(r(\theta_i)^2 + 2\,r'(\theta_i)^2\right)\Delta \theta_i^2}{A},
\end{equation}
where $n$ is the total number of cells.
Since Eq.~\eqref{eq:r_c} is in a discrete form, we approach the problem of finding the global minimum in the continuum limit using the Cauchy--Schwarz inequality, rather than relying on derivatives.
By the Cauchy--Schwarz inequality,
\begin{equation}
\label{eq:cs_inequal}
\begin{aligned}
\sum_{i=1}^{n} &\frac{1}{16}\, r(\theta_i)\left|\kappa(\theta_i)\right|
\left(r(\theta_i)^2 + 2\,r'(\theta_i)^2\right)\Delta \theta_i^2 \cdot (1^2 + \cdots + 1^2) \\
&\ge
\left(\sum_{i=1}^{n} \sqrt{\frac{1}{16}\, r(\theta_i)\left|\kappa(\theta_i)\right|
\left(r(\theta_i)^2 + 2\,r'(\theta_i)^2\right)}\, \Delta \theta_i \right)^2
\end{aligned}
\end{equation}
and thus
\begin{equation}
\label{eq:cr_limit}
r_c \ge \frac{\left(\sum_{i=1}^{n} \sqrt{\frac{1}{16}\, r(\theta_i)\left|\kappa(\theta_i)\right|\left(r(\theta_i)^2 + 2\,r'(\theta_i)^2\right)}\, \Delta \theta_i \right)^2}{A} \cdot \frac{1}{n}.
\end{equation}
Equality holds in the Cauchy--Schwarz inequality when all cells have equal cross-sectional areas, i.e., $a_1 = \cdots = a_i = \cdots = a_n$, where $a_i = \frac{1}{16}\, r(\theta_i)\left|\kappa(\theta_i)\right|\left(r(\theta_i)^2 + 2\,r'(\theta_i)^2\right)\Delta \theta_i^2$, and the corresponding $r_c$ attains the lower limit.
Therefore, the optimized contraction ratio $r_{c|\mathrm{opt}}$ is given by
\begin{equation}
\label{eq:rc_opt}
r_{c|\mathrm{opt}} = \frac{\left( \int_{0}^{2\pi} \sqrt{\frac{1}{16}\, r(\theta)\left|\kappa(\theta)\right|\left(r(\theta)^2 + 2\,r'(\theta)^2\right)} \, d\theta \right)^2}{A} \cdot \frac{1}{n}.
\end{equation}
Note that Eq.~\eqref{eq:rc_opt} approaches the global minimum in the continuum limit($\Delta \theta_i$ is very small), and the summation term $\sum_{i=1}^{n} \sqrt{\frac{1}{16}\, r(\theta_i)\left|\kappa(\theta_i)\right|\left(r(\theta_i)^2 + 2\,r'(\theta_i)^2\right)}\, \Delta \theta_i$ is written in integral form because $\Delta \theta_i$ is sufficiently small.

\subsection{Design and fabrication of a UAV for flight testing}
\subsubsection{Aerodynamic design}
The objective of the aircraft design was to develop a demonstrator aircraft capable of stable flight with the origami wing prototype described in the main text. 
The aircraft-level design was therefore constrained by the geometric and mass properties of the wing prototype. 
Specifically, the initial target maximum takeoff weight (MTOW) was set to 750 g, while the origami wing had a half-span of 0.6 m, a chord length of 0.25 m, and a NACA 2412 airfoil. 
Including the center wing mount, the total wingspan of the aircraft was 1.26 m, and the mass of each half-wing was limited to 250 g.
In addition, the fabricated origami wing did not readily allow the integration of conventional ailerons.
Therefore, a dihedral angle of approximately $10^\circ$ was incorporated into the center wing mount to provide passive roll stability, and roll control was achieved through rudder input.

The overall aircraft configuration was designed using the Athena Vortex Lattice (AVL) method~\cite{Drela_AVL}.
The incidence angles of the wing and horizontal stabilizer were adjusted such that the aircraft trimmed near zero angle of attack under nominal cruise conditions.
Tail sizing was additionally guided by conventional horizontal and vertical tail volume coefficient considerations.
The resulting aircraft geometry and aerodynamic specifications are summarized in Table~\ref{tab:aircraft_geometry}.

AVL analysis was used to estimate the aerodynamic neutral point and determine the center-of-gravity (CG) location corresponding to the target static margin.
Because the neutral point location is not highly sensitive to airfoil shape or incidence angle, the neutral point position was calculated in AVL using Eq.~\eqref{eq:xNP} measured from the root leading edge of the wing.
\begin{equation}
\label{eq:xNP}
x_{NP} = 0.132.
\end{equation}
The aircraft CG location was subsequently determined using the desired static margin of 12.8\%, as shown in Eq.~\eqref{eq:xCG},
\begin{equation}
\label{eq:xCG}
x_{CG} = x_{NP} - SM \cdot \bar{c} = 0.10,
\end{equation}
where $SM$ is the static margin and $\bar{c}$ is the mean aerodynamic chord.

The aerodynamic drag characteristics of the aircraft were estimated using a component-based drag buildup method.
The overall drag polar was modeled as
\begin{equation}
\label{eq:overall_drag_polar}
C_D = C_{D_p} + k\left(C_L - C_{L,\min}\right)^2
\end{equation}
where $C_{D_p}$ represents the lift-independent parasite drag contribution and the second term represents the lift-dependent drag. 
Note that second quadratic term is the sum of lift-dependent parasite drag contribution and the induced drag.

The profile drag of the wing and horizontal stabilizer was estimated by combining sectional lift coefficients obtained from AVL with sectional drag polars generated using XFOIL at the corresponding Reynolds numbers.
The Reynolds numbers of the main wing, horizontal stabilizer, and vertical stabilizer were assumed to be 130{,}310, 91{,}910, and 81{,}000, respectively.
The corresponding drag polars obtained using XFOIL are shown in Fig.~\ref{fig:drag_polar}.
The sectional drag contributions were integrated spanwise to obtain the total profile drag.
This procedure was repeated at multiple aircraft lift coefficients to characterize the lift-dependent profile drag behavior (See Table~\ref{tab:profile_drag_coefficients}).
The induced drag component was directly obtained from AVL analysis.

The fuselage drag was estimated using a skin-friction-based drag model with an appropriate form-factor correction.
Additional drag contributions associated with the front flat area and rear base area were also included.
Because the aircraft did not employ landing gear, exposed antennas, or other major external protrusions, additional excrescence drag contributions were assumed to be negligible.
Table~\ref{tab:fuselage_drag_buildup} summarizes the fuselage drag buildup.
Table~\ref{tab:aircraft_drag} summarizes all drag components expressed as lift-independent and lift-dependent contributions. Note that the drag of the vertical stabilizer was estimated using the zero-lift drag coefficient of the NACA 0012 airfoil and the $C_{Ds}$ are based on the reference wing area.

The propulsion system was designed based on the target MTOW using electric propulsion performance analysis methods.
Motor and propeller combinations were selected to provide sufficient thrust margin while maintaining acceptable cruise efficiency and takeoff performance. Because of the low aircraft mass, the smallest motor series available from T-Motor was selected as the design baseline.
Three motor $K_v$ values and three APC propellers were evaluated, resulting in a total of nine motor-propeller combinations.
Table~\ref{tab:propulsion_efficiencies} summarizes the overall propulsion efficiencies for the evaluated combinations.
The highest-efficiency configuration employed the AS2304 Kv1800 motor with the APC Sport 9×6 propeller, resulting in an overall efficiency of 45.4\%.

\subsubsection{Fuselage construction}
The fuselage, horizontal stabilizer, and vertical stabilizer were fabricated from balsa wood according to the aircraft design described above.
The wing mount was designed to provide a dihedral angle of $10^\circ$, as shown in Fig.~\ref{fig:aircraft_fabricated}A.
It was fabricated from plywood to provide sufficient local stiffness while maintaining a low structural mass.

During the final integration process, the high mass fraction of the deployable origami wing significantly constrained the achievable CG location.
To resolve this issue, the design static margin was reduced from 12.8\% to approximately 10.8\%, and additional ballast mass was added near the nose section to shift the CG forward.
As a result, the final aircraft mass increased from the initial 750 g target to approximately 850 g. 
In addition, propeller was changed from APC sports 9x6 to GWS 9050 that is equivalent to 9x5 as the APC propeller was not compatible with the propeller mount of the motor.
Fig.~\ref{fig:aircraft_fabricated}B shows the finished demonstrator with the wings fully deployed.
The detailed specifications of the completed demonstrator aircraft are summarized in Table~\ref{tab:final_aircraft_specs}.

\subsubsection{Pre-flight loading test}
Prior to the flight test, we conducted a preliminary loading test to verify the load-bearing performance of the fabricated origami wing.
As each wing weighs 260~g and the fuselage weighs approximately 250~g (total approximately 750~g), and a safety factor of 2 is adopted, we assume that a lift equivalent to 750~g acts on a single wing.
Additionally, we assume that the lift is elliptically distributed and apply it by converting the distributed load into equivalent lumped loads (see Fig.~\ref{fig:loading_test}A).
As shown in Fig.~\ref{fig:loading_test}B, we apply the lumped loads using water bottles, and the wing withstands a load corresponding to 750~g (three times the weight of a single wing).

\clearpage


\clearpage

\newgeometry{left=0.75in,right=0.75in,top=1in,bottom=1in}

\begin{figure*}[ht]
    \centering
    \includegraphics[width=178mm]{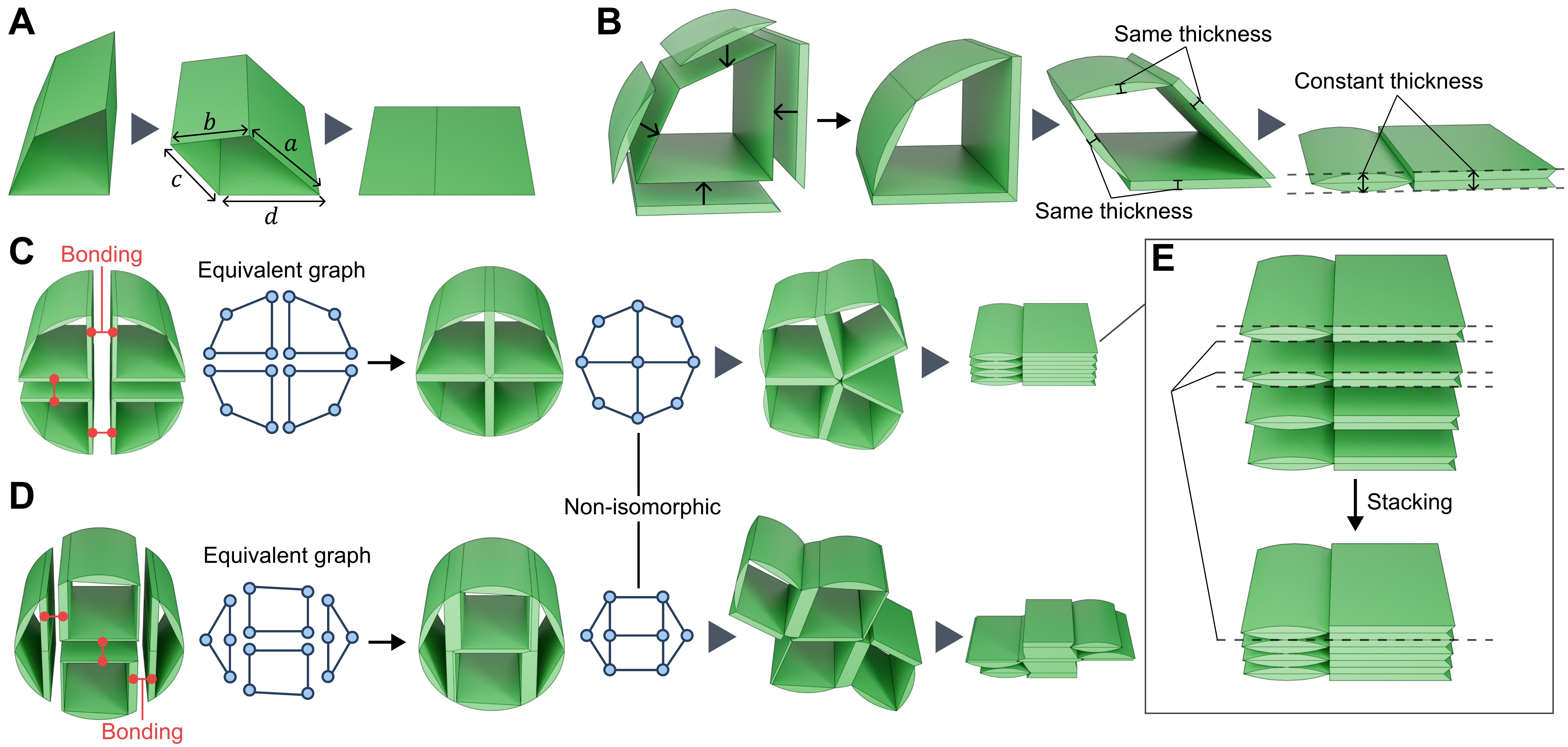}
  \caption{\textbf{Thickness-dependent flat-foldability constraint.} (\textbf{A}) Flat-foldable prismatic tube. (\textbf{B}) Corresponding flat-foldable volumetric origami tube with constant panel thickness. (\textbf{C}) Flat-foldable volumetric origami targeting a cylinder. (\textbf{D}) Flat-foldable volumetric origami targeting a cylinder with a different origami layout. (\textbf{E}) Exploded view of the flat-folded configuration in (C).}
    \label{fig:thicknesswise_flat_foldable}
\end{figure*}

\begin{figure*}[ht]
    \centering
    \includegraphics[width=178mm]{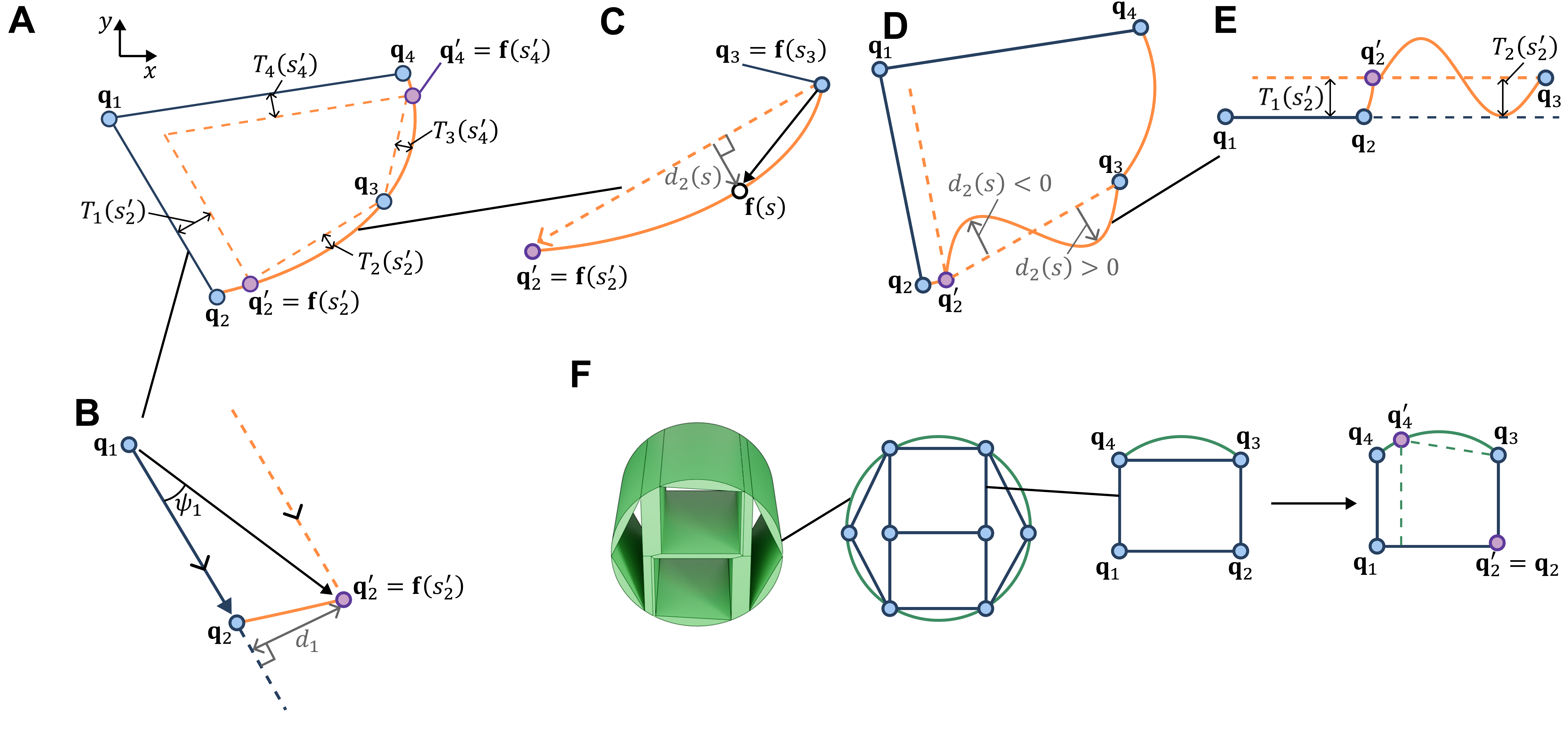}
  \caption{\textbf{Geometric definitions and exceptional cases for deriving the thickness-dependent flat-foldability constraints.} (\textbf{A}) Definitions of $T_1(s_2')$, $T_2(s_2')$, $T_3(s_4')$, and $T_4(s_4')$. (\textbf{B}) Local geometric definitions at $\mathbf{q}_2$. (\textbf{C}) Local geometric definitions at $\mathbf{q}_3$. (\textbf{D}) A case where $\mathbf{f}(s)$ contains a concave region. (\textbf{E}) Configuration corresponding to a $180^\circ$ folding angle at $\mathbf{q}_2'$ in (D). (\textbf{F}) Exceptional case where the edge $\mathbf{q}_2\mathbf{q}_3$ is an internal edge. }
    \label{fig:cylinder_thickness}
\end{figure*}

\begin{figure*}[ht]
    \centering
    \includegraphics[width=178mm]{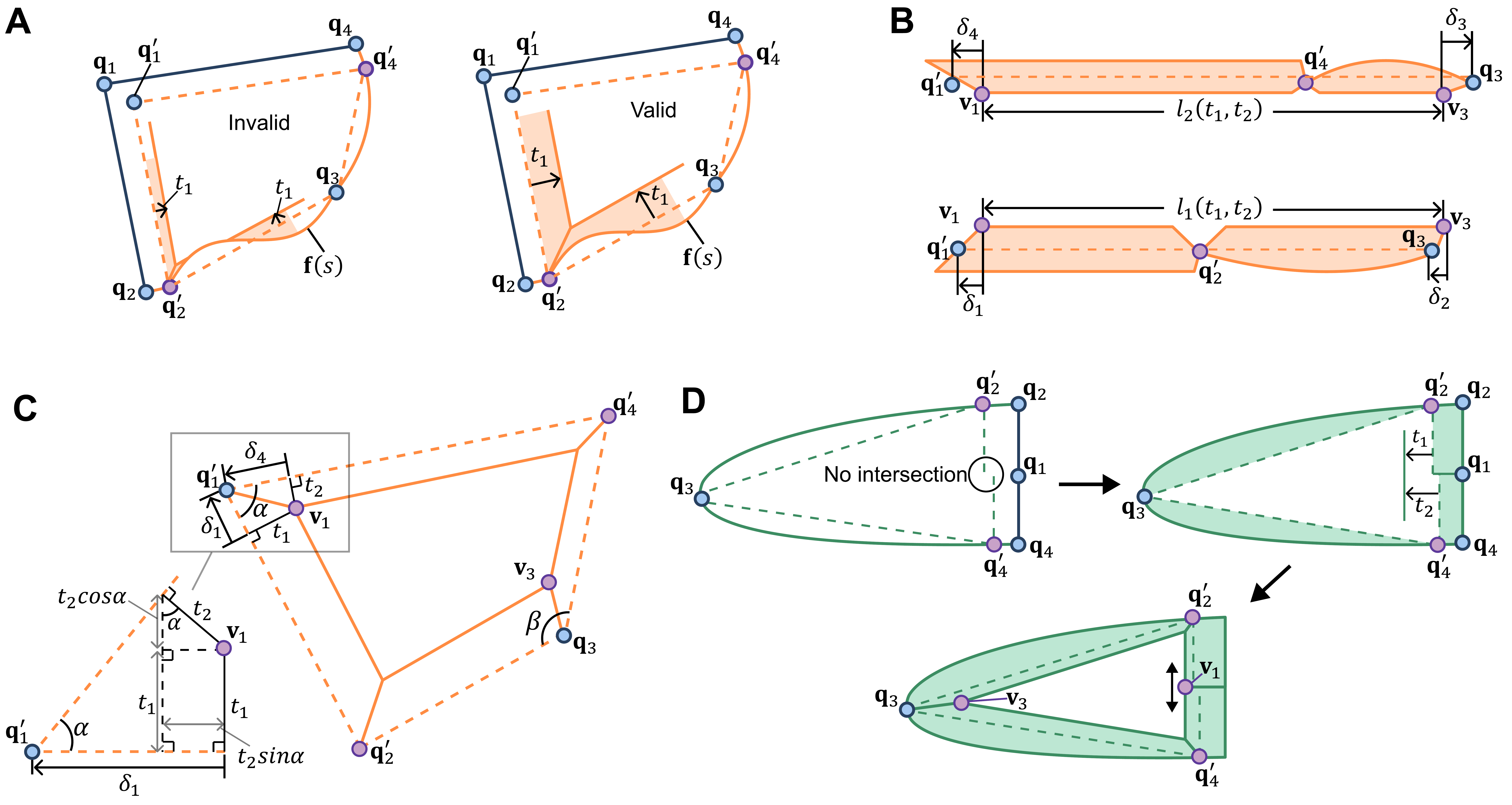}
  \caption{\textbf{Geometric definitions and exceptional cases for deriving the lengthwise flat-foldability constraints.} (\textbf{A}) Admissible offset distance when inward indentation exists. (\textbf{B}) Geometric definitions of $\delta_1$–$\delta_4$. (\textbf{C}) Detailed local geometry around $\mathbf{q}_1'$. (\textbf{D}) Exceptional case where $\mathbf{q}_1$, $\mathbf{q}_2$, and $\mathbf{q}_4$ are collinear and $\mathbf{q}_1'$ is not well-defined. }
    \label{fig:cylinder_length}
\end{figure*}

\begin{figure*}[ht]
    \centering
    \includegraphics[width=178mm]{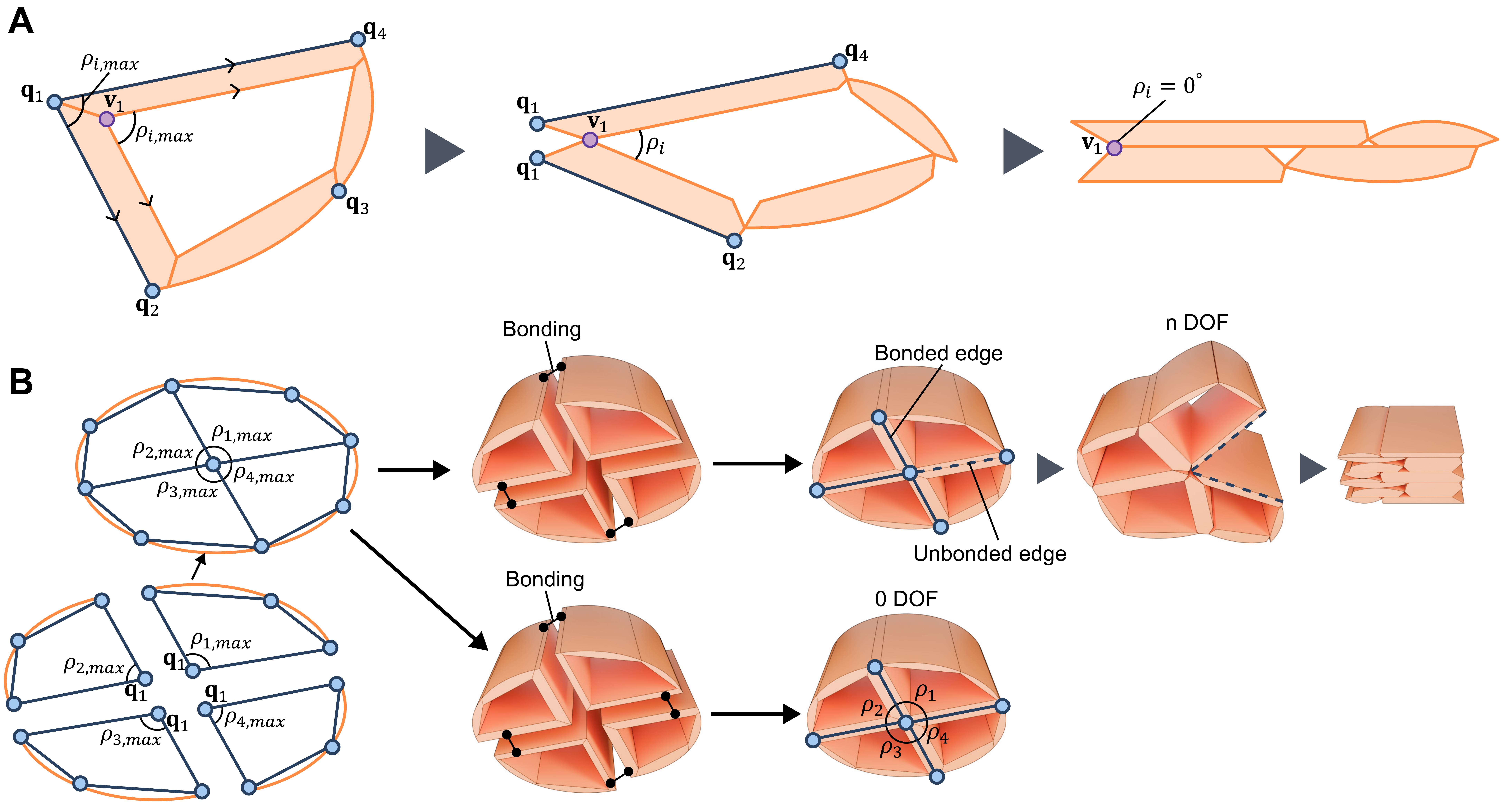}
  \caption{\textbf{Definition of folding angle and degree of freedom under full and partial bonding at internal edges.} (\textbf{A}) Folding motion of the flat-foldable thick origami cell and the definition of the folding angle $\rho_i$ at $\mathbf{v}_1$.  (\textbf{B}) Comparison of the degree of freedom between the case where bonding is applied at all internal edges except one (top) and the case where bonding is applied at all internal edges (bottom).  }
    \label{fig:locking}
\end{figure*}

\begin{figure*}[ht]
    \centering
    \includegraphics[width=178mm]{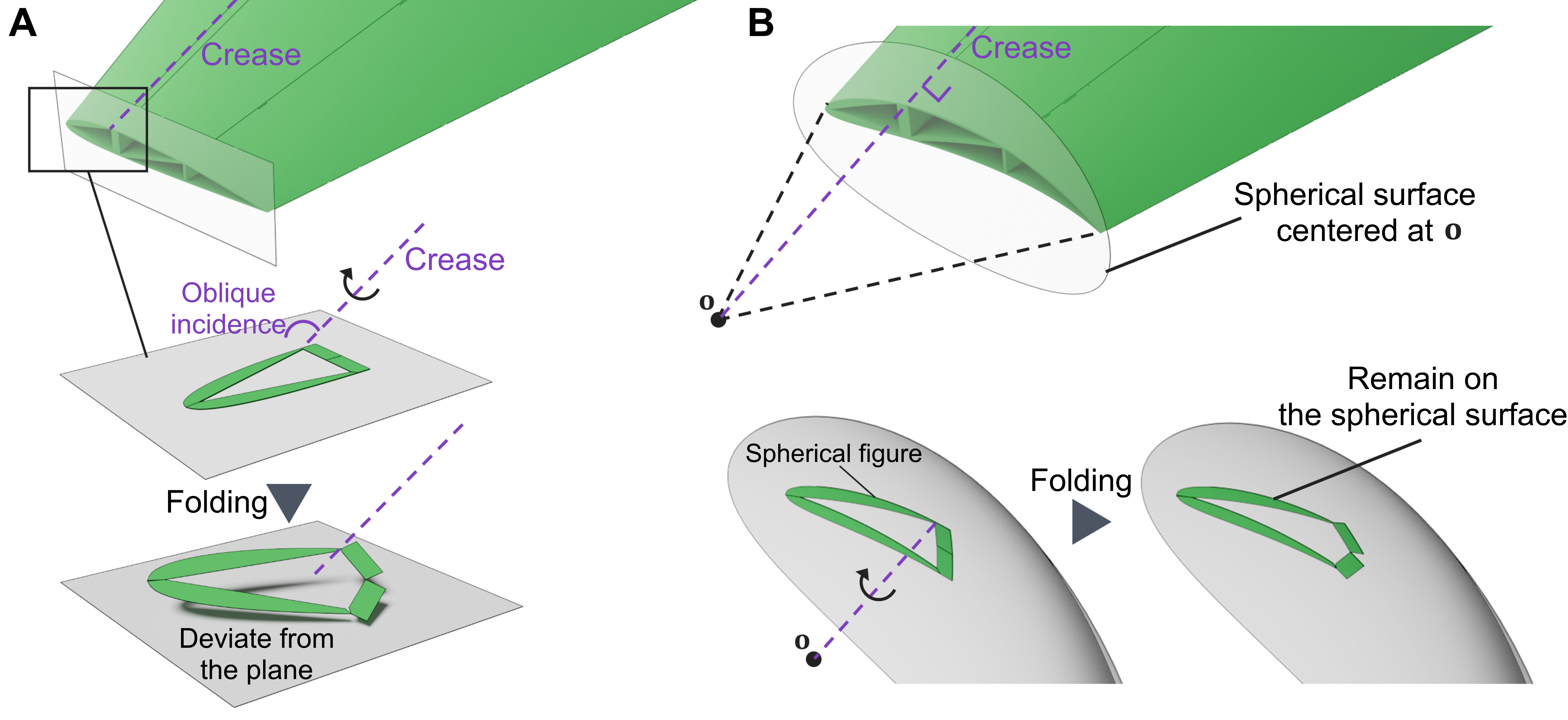}
  \caption{\textbf{Planar versus spherical representations of volumetric origami cells for generalized cones.} (\textbf{A}) Folding motion of a volumetric origami cell for a generalized cone, represented on a plane. A volumetric origami cell represented on a plane deviates from the plane during folding due to creases that pass through the plane obliquely. (\textbf{B}) Folding motion of a volumetric origami cell for a generalized cone, represented on a spherical surface. A volumetric origami cell represented on a sphere remains on the sphere during folding, as the creases are always normal to the sphere centered at $\mathbf{o}$.  }
    \label{fig:cone_problem}
\end{figure*}

\begin{figure*}[ht]
    \centering
    \includegraphics[width=178mm]{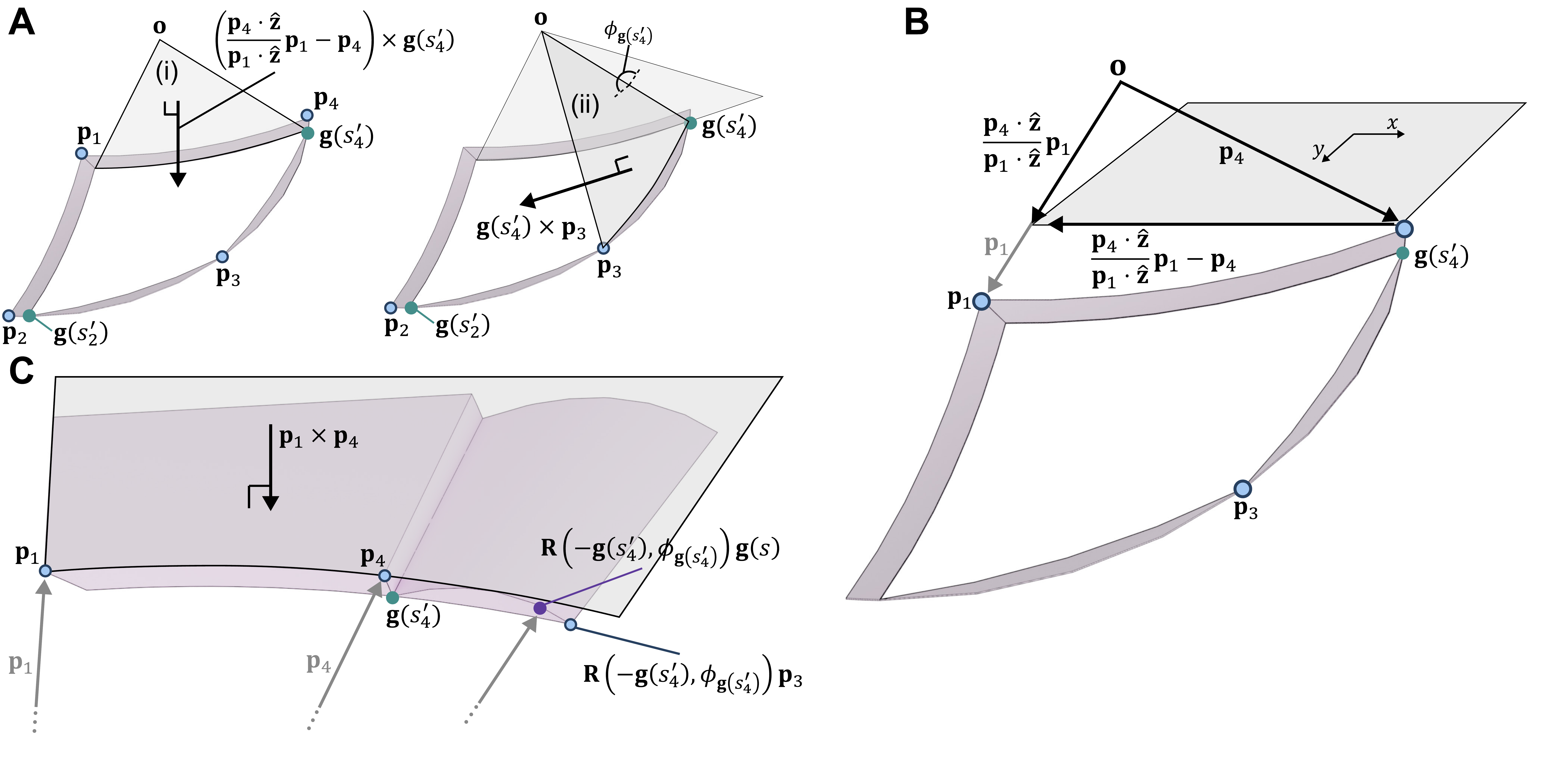}
  \caption{\textbf{Geometric illustration of thickness-dependent flat-foldability constraints for a generalized cone.} (\textbf{A}) Definition of internal facets determined by the great-circle arcs. (\textbf{B}) Geometric illustration of the rotation axis $\frac{\mathbf{p}_4 \cdot \hat{\mathbf{z}}}{\mathbf{p}_1 \cdot \hat{\mathbf{z}}}\,\mathbf{p}_1 - \mathbf{p}_4$. (\textbf{C}) A portion of the volumetric origami cell in the flat-folded state is bounded by the plane defined by $\mathbf{p}_1$ and $\mathbf{p}_4$.}
    \label{fig:cone_thickness}
\end{figure*}

\begin{figure*}[ht]
    \centering
    \includegraphics[width=178mm]{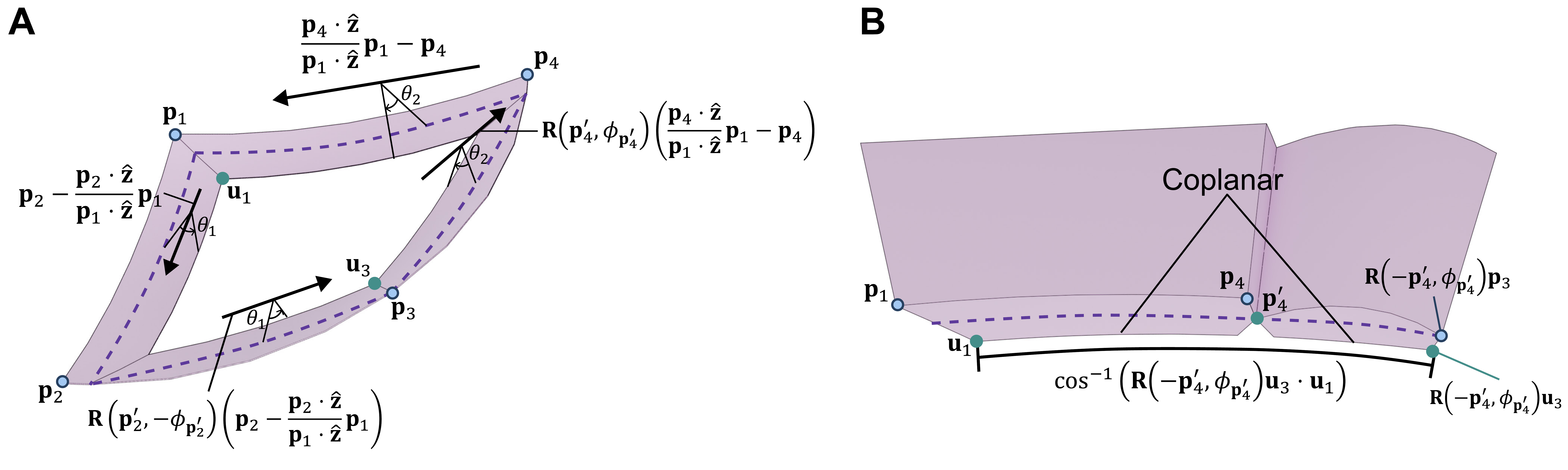}
  \caption{\textbf{Geometric illustration of lengthwise flat-foldability constraints for a generalized cone.} (\textbf{A}) Rotation axes and angular adjustments used to determine the internal facets. (\textbf{B}) Spherical distance between creases $\mathbf{u}_1$ and $\mathbf{u}_3$ in the flat-folded state.}
    \label{fig:cone_length}
\end{figure*}

\begin{figure*}[ht]
    \centering
    \includegraphics[width=178mm]{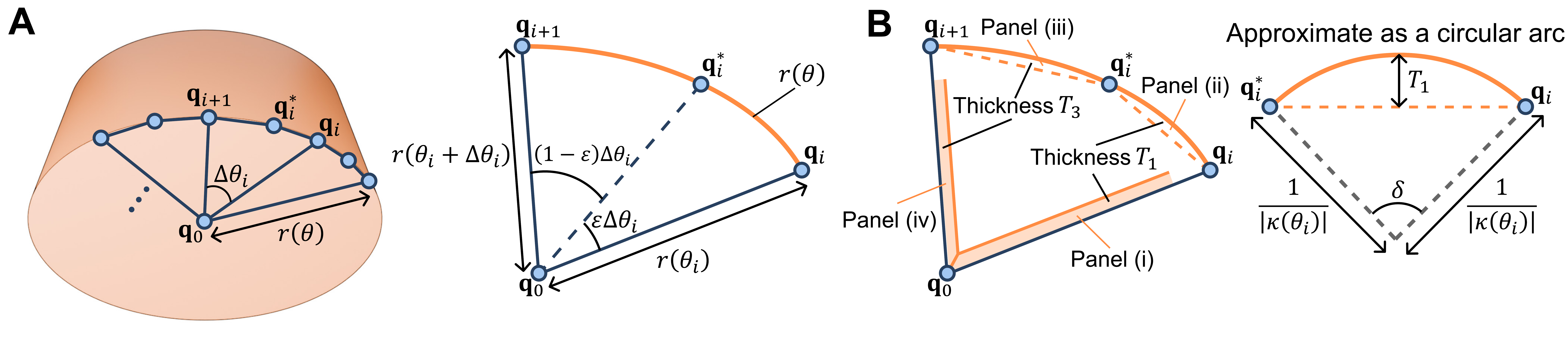}
  \caption{\textbf{Geometric definition for the derivation of the optimal contraction ratio in the continuum limit.} (\textbf{A}) Geometric definitions of an infinitesimal cell used in the continuum-limit derivation of the contraction ratio. (\textbf{B}) Circular arc approximation after determining the position of $\mathbf{q}_i^*$.   }
    \label{fig:analysis_geometry}
\end{figure*}

\begin{figure*}[ht]
    \centering
    \includegraphics[width=178mm]{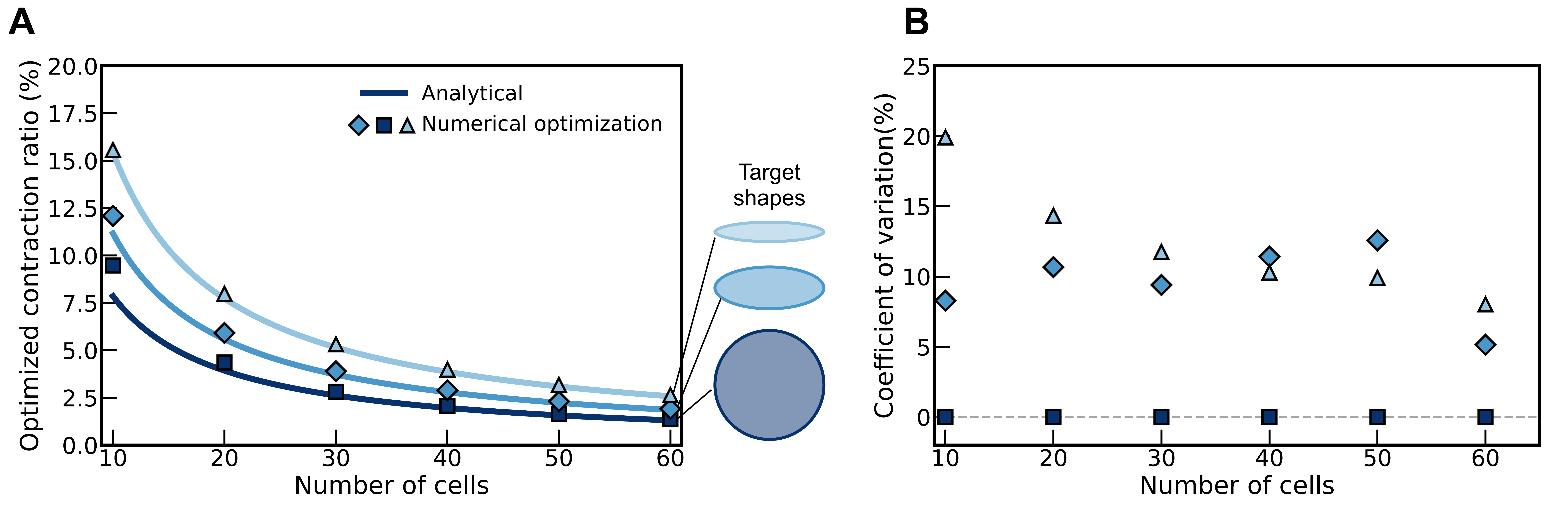}
  \caption{\textbf{Optimal contraction ratio versus the number of cells (large-cell limit).} (\textbf{A}) Plot of the optimized contraction ratio versus the number of cells $n$ for generalized cylinders with elliptical cross-sections of aspect ratios $1$, $0.4$, and $0.2$. The solid lines represent the analytical continuum-limit expression of the optimal contraction ratio, while the markers indicate local minima obtained through numerical optimization. (\textbf{B}) Plot of the coefficient of variation of the origami cell bounded areas $\{a_1, a_2, \ldots, a_n\}$ obtained from each numerical optimization.  }
    \label{fig:analysis_limit}
\end{figure*}

\begin{figure*}[ht]
    \centering
    \includegraphics[width=178mm]{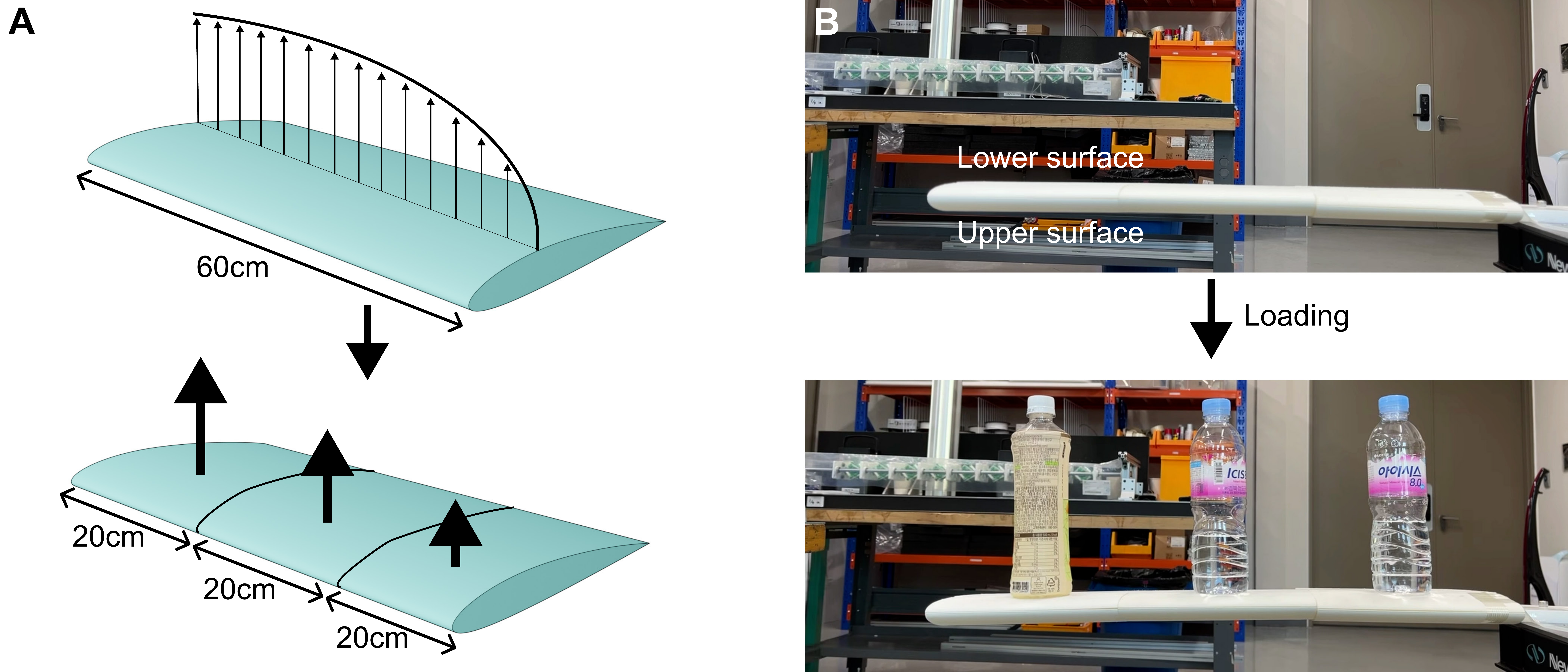}
  \caption{\textbf{Pre-flight simple loading test.} (\textbf{A}) Continuous, elliptically distributed lift is converted into discrete lumped loads. (\textbf{B}) Images of the origami wing before and after load application.  }
    \label{fig:loading_test}
\end{figure*}

\begin{figure*}[ht]
    \centering
    \includegraphics[width=178mm]{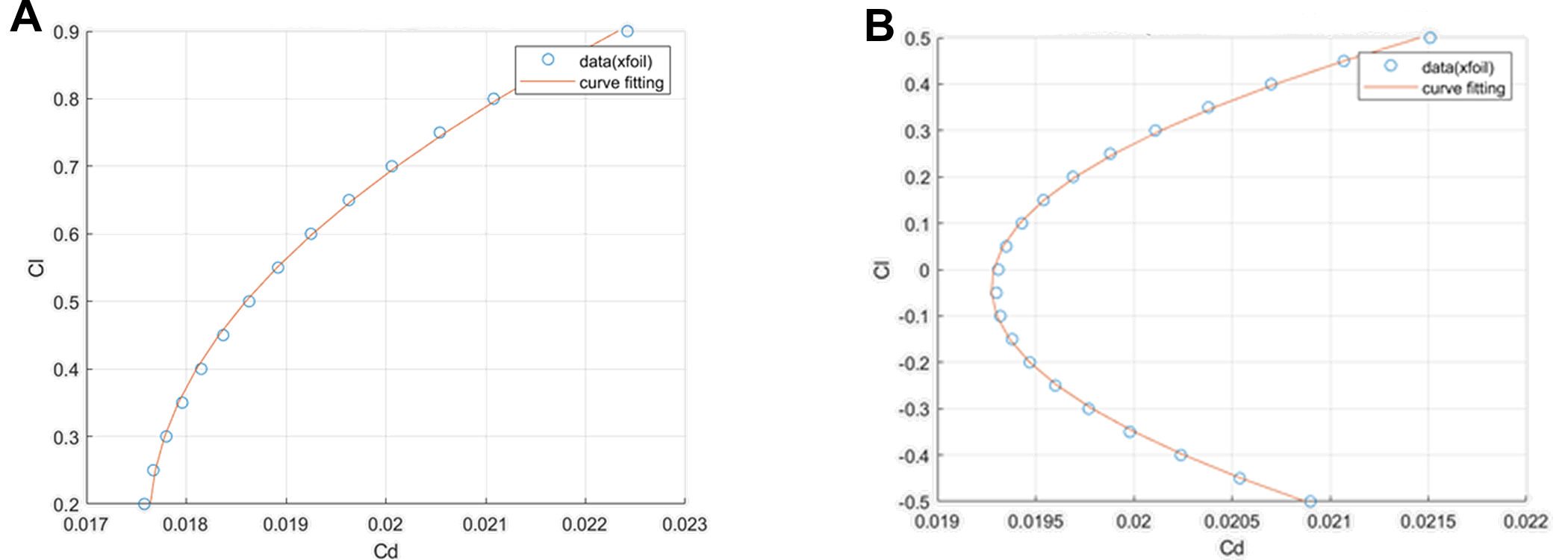}
  \caption{\textbf{Drag polars of the main wing and horizontal stabilizer.} (\textbf{A}) Blue dots indicate the drag polar obtained using XFOIL for the main wing (NACA 2412 airfoil) at a Reynolds number of 130{,}310, and the orange curve represents the corresponding quadratic curve fit. (\textbf{B})  Blue dots indicate the drag polar obtained using XFOIL for the horizontal stabilizer (NACA 0012 airfoil) at a Reynolds number of 91{,}910, and the orange curve represents the corresponding quadratic curve fit.  }
    \label{fig:drag_polar}
\end{figure*}

\begin{figure*}[ht]
    \centering
    \includegraphics[width=178mm]{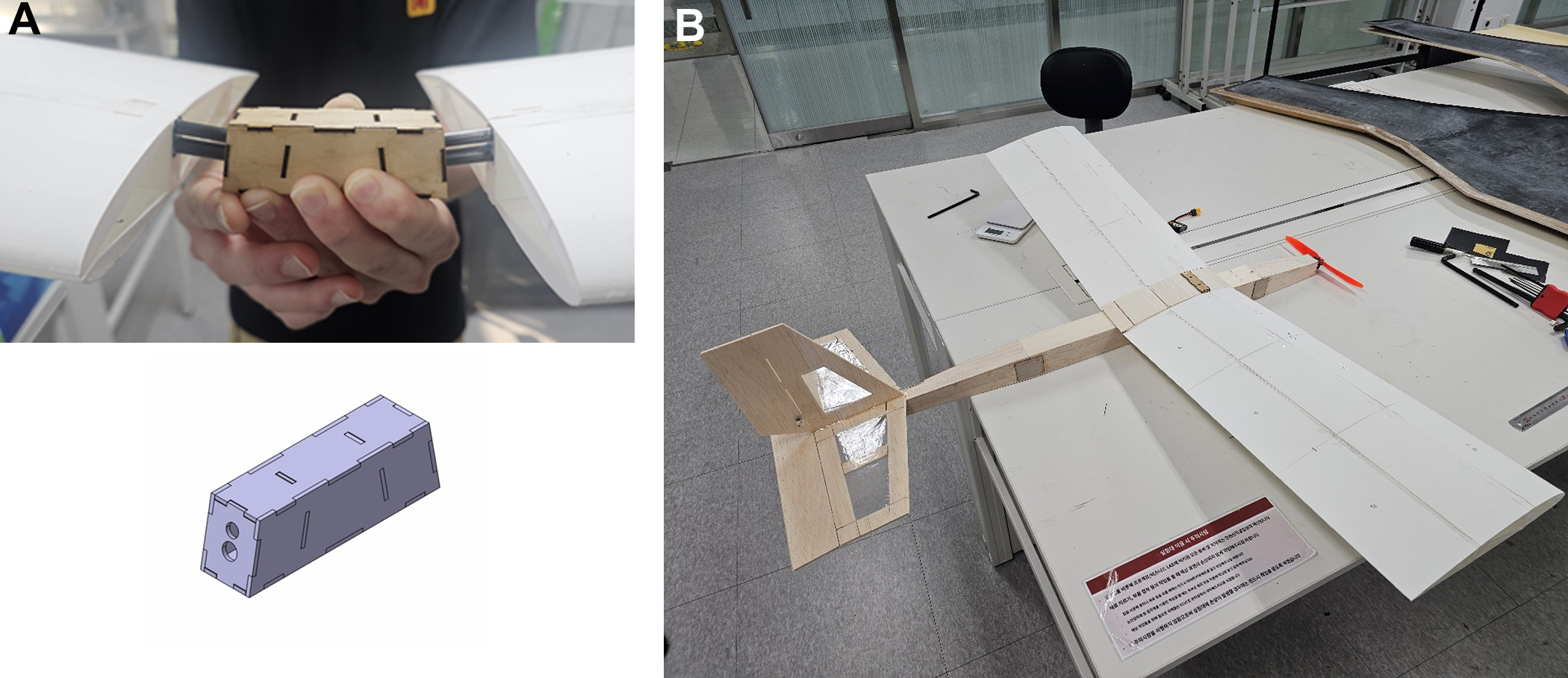}
  \caption{\textbf{Fabricated demonstrator aircraft.} (\textbf{A})  Fabricated wing mount and corresponding CAD model. (\textbf{B}) Final demonstrator aircraft prototype equipped with the deployable origami wing.  }
    \label{fig:aircraft_fabricated}
\end{figure*}

\clearpage

\begin{table}[ht]
\centering
\caption{Geometric and aerodynamic specifications of the wing and tail surfaces.}
\label{tab:aircraft_geometry}
\small
\begin{tabular}{p{0.38\linewidth}ccc}
\hline
Parameter & Wing & Horizontal stabilizer & Vertical stabilizer \\
\hline
Area, $S$ ($\mathrm{m}^2$) & 0.315 & 0.077 & 0.030 \\
Aspect ratio, $\mathcal{A}$ & 5.04 & 2.51 & 1.33 \\
Sweep, $\Lambda$ & $0^\circ$ & $12.8^\circ$ & $26.6^\circ$ \\
Airfoil & NACA 2412 & NACA 0012 & NACA 0012 \\
Taper ratio, $\lambda$ & 1.00 & 0.75 & 0.50 \\
Incidence angle & $6.1^\circ$ & $4.9^\circ$ & $0^\circ$ \\
\hline
Distance between root LEs of wing and horizontal stabilizers & \multicolumn{3}{c}{0.7 m} \\
Aircraft mass & \multicolumn{3}{c}{0.75 kg} \\
\hline
\end{tabular}
\end{table}

\begin{table}[ht]
\centering
\caption{Combined profile drag coefficients of the main wing and horizontal stabilizer at different aircraft lift coefficients.}
\label{tab:profile_drag_coefficients}
\begin{tabular}{cc}
\hline
Aircraft $C_L$ & Combined profile $C_D$ \\
\hline
0.42 & 0.0230 \\
0.62 & 0.0241 \\
0.82 & 0.0260 \\
\hline
\end{tabular}
\end{table}

\begin{table}[ht]
\centering
\caption{Component-based fuselage drag buildup estimation.}
\label{tab:fuselage_drag_buildup}
\begin{tabular}{lc}
\hline
Parameter & Value \\
\hline
Reynolds number & 517{,}000 \\
Skin friction coefficient (fully turbulent) & 0.0051 \\
Fineness ratio & 20 \\
Form factor & 1.057 \\
Wetted area & $0.135~\mathrm{m}^2$ \\
$C_D$ from skin friction & 0.0023 \\
$C_D$ from the front flat area & 0.0022 \\
$C_D$ from the rear base area & 0.0005 \\
\hline
\end{tabular}
\end{table}

\begin{table}[ht]
\centering
\caption{Summary of the aircraft drag model including lift-independent and lift-dependent contributions.}
\label{tab:aircraft_drag}
\begin{tabular}{lc}
\hline
Component & $C_D$ based on the reference wing area \\
\hline
Main wing and horizontal stabilizer 
& $0.010C_L^2 - 0.0049C_L + 0.0233$ \\
& $(0.42 \leq C_L \leq 0.82)$ \\
Vertical stabilizer & 0.0022 \\
Fuselage & 0.0050 \\
Induced drag & $\dfrac{1}{\pi e\mathcal{A}}C_L^2 = 0.062C_L^2$ \\
\hline
Total & $C_D = 0.0286 + 0.072\left(C_L - 0.0341\right)^2$ \\
\hline
\end{tabular}
\end{table}

\begin{table}[ht]
\centering
\caption{Overall propulsion efficiencies for the evaluated motor--propeller combinations.}
\label{tab:propulsion_efficiencies}
\begin{tabular}{lccc}
\hline
Motor & APC SP 7$\times$6 & APC SP 8$\times$6 & APC SP 9$\times$6 \\
\hline
AS2304 Kv 1500 & $\eta = 43.2\%$ & $\eta = 42.1\%$ & $\eta = 44.8\%$ \\
AS2304 Kv 1800 & $\eta = 43.3\%$ & $\eta = 42.4\%$ & $\eta = 45.4\%$ \\
AS2304 Kv 2300 & $\eta = 42.0\%$ & $\eta = 40.9\%$ & $\eta = 43.8\%$ \\
\hline
\end{tabular}
\end{table}

\begin{table}[ht]
\centering
\caption{Final specifications of the demonstrator aircraft.}
\label{tab:final_aircraft_specs}
\begin{tabular}{lclc}
\hline
Parameter & Value & Parameter & Value \\
\hline
Reference wing area & $0.30~\mathrm{m}^2$ 
& CG location from root LE & $0.100~\mathrm{m}$ \\

Mass & $0.85~\mathrm{kg}$ 
& NP location from root LE & $0.127~\mathrm{m}$ \\

Static margin & $10.8\%$ 
& Distance between root LEs & $0.62~\mathrm{m}$ \\

Trim $C_L$ & 0.65 
& Tail volume coefficient (H) & 0.62 \\

Cruise speed & $8.5~\mathrm{m/s}$ 
& Tail volume coefficient (V) & 0.05 \\
\hline
\end{tabular}
\end{table}


\clearpage 

\subsection*{List of Supplementary Movies}

\paragraph{Caption for Movie~S1.}
\textbf{Folding motion of the volumetric origami prototypes}
Folding motion of volumetric origami prototypes targeting generalized cylinders with clover- and heart-shaped cross-sectional profiles and generalized cones with circular and elliptic cross-sectional profiles.

\paragraph{Caption for Movie~S2.}
\textbf{Folding motion of the origami wing and flight test}
Deployment process of a flat-foldable origami wing and subsequent flight testing using a custom UAV equipped with the origami wing.


\end{document}